\documentclass[apj]{emulateapj}
\usepackage{graphicx}
\usepackage{amsmath}
\usepackage[caption=false]{subfig}
\usepackage{natbib}

\begin{document}

\title{Globally Distributed Energetic Neutral Atom Maps for the ``Croissant" Heliosphere}

\author{M. Kornbleuth\altaffilmark{1}}
\affil{\altaffilmark{1}Astronomy Department, Boston University, Boston, MA 02215, USA} 
\email{kmarc@bu.edu}

\author{M. Opher\altaffilmark{1}\altaffilmark{,}\altaffilmark{2}}
\affil{\altaffilmark{2}Harvard-Smithsonian Center for Astrophysics, 60 Garden St. Cambridge, MA, USA }

\author{A. T. Michael\altaffilmark{1}}

\author{J. F. Drake\altaffilmark{3}}
\affil{\altaffilmark{3}Department of Physics and the Institute for Physical Science and Technology, University of Maryland, College Park, MD, USA} 

\begin{abstract}
A recent study by \cite{Opher15} suggested the heliosphere has a ``croissant" shape, where the heliosheath plasma 
is confined by the toroidal solar magnetic field. The ``croissant" heliosphere is in contrast to the 
classically accepted view of a comet-like tail. We investigate the effect of the ``croissant" heliosphere model on 
energetic neutral atom (ENA) maps. Regardless of the existence of a split tail, the confinement of the heliosheath plasma 
should appear in ENA maps. ENA maps from the Interstellar Boundary Explorer (IBEX) have shown two high latitude 
lobes with excess ENA flux at higher energies in the tail of the heliosphere. These lobes could be a signature of the 
confinement of the heliosheath plasma, while some have argued they are caused by the fast/slow solar wind profile. Here we
present ENA maps of the ``croissant" heliosphere, focusing on understanding the effect of the heliosheath plasma collimation by the solar magnetic field while using a uniform solar wind. We incorporate pick-up ions (PUIs) into 
our model based on \cite{Malama06} and \cite{Zank10}. We use the neutral solution from our MHD model to determine the angular variation of the PUIs, and include the extinction of PUIs in the heliosheath. In the presence of a uniform solar wind, we find that the collimation in the ``croissant" heliosphere does manifest itself into two high latitude lobes of increased ENA flux in the downwind direction. 
\end{abstract}

\keywords{ISM: atoms - magnetohydrodynamics (MHD) - solar wind - Sun: heliosphere} 

\section{Introduction}
The interaction between the solar wind and the partially ionized gas of the local interstellar medium (LISM) creates a 
bubble known as the heliosphere. Classically, the shape of the heliosphere has been regarded as comet-like, with a long tail pointed in the direction opposite the Sun's motion through the LISM \citep{Parker61, Baranov93}. 
In this view, the solar magnetic field was assumed to have a negligible effect on the global structure of the 
heliosphere. Computational models based on the magnetohydrodynamic (MHD) equations which include the LISM and a dipolar solar wind magnetic field have been able to reproduce this long tail structure \citep{Opher06, Pogorelov07, Opher09, Washimi11, Pogorelov13, OpherDrake13}. However, \cite{Opher15} and \cite{Drake15} used a unipolar solar magnetic field to limit artificial magnetic dissipation inherent in modeling the current sheet. They argued that the magnetic tension of the solar wind magnetic field instead alters the flows in the heliosheath and produces a ``croissant" heliosphere. While the inclusion of a dipolar solar magnetic field can weaken this effect, the use of this magnetic field configuration does not necessarily lead to a long tail structure \citep{Opher16, Michael18}. 

The \cite{Opher15} model is based on a 3D MHD simulation of a single ion fluid with four neutral fluids, and  displays a shortened heliotail due to the presence of the lobes. The interstellar plasma is able to flow between these two jets, where a pressure balance between the thermal pressure of the interstellar plasma with the magnetic and thermal pressure of the lobes occurs. The jets were shown to exhibit turbulence, with turnover timescales on the order of years for the largest eddies. \cite{Drake15} confirmed the existence of the jets with an axisymmetric analytic model of the heliosheath. Using data from the Ion and Neutral Camera (INCA) on board the Cassini spacecraft and Voyager data, \cite{Dialynas17}  argue that the heliosphere is tail-less, being ``bubble-like". While the ``bubble-like" shape is an idealization as the solar wind plasma needs to be able to escape, it does agree with the model of \cite{Opher15} in arguing for a tail-less heliosphere. It is important to emphasize that the two models do differ in the mechanism which structures the heliosphere. \cite{Dialynas17} suggest that the ``bubble-like" heliosphere is created by a strong interstellar magnetic field (as in \cite{Parker61}), while \cite{Opher15} show a structure and argue (see model by \cite{Drake15}) that the solar magnetic field plays a fundamental role in organizing the heliosphere. While both models suggest a shortened tail, these different structures would likely produce different observational signatures in ENAs.

\cite{Pogorelov15} claimed that modeling the neutrals kinetically, as opposed to the four fluid approximation, would cause the disappearance of the two tails since the fluid-like treatment of the neutrals would suppress charge exchange across the region separating the lobes. \cite{Alexashov05} studied the effect of kinetic neutrals as compared to multi-fluid neutrals and found the plasma solutions to be similar within $\sim$5\% in the nose direction. Charge exchange strongly influences the heliosphere; however, the findings by \cite{Alexashov05} suggest that the inclusion of kinetic neutrals would not be sufficient to remove the jets in \cite{Opher15}. This is seen in \cite{Izmodenov15} where evidence of collimation of the heliosheath plasma by the solar magnetic field is seen even with neutrals modeled kinetically. Other possible effects that could suppress the two tails are the inclusion of solar cycle and the numerical erosion of the solar magnetic field between the two tails \citep{Pogorelov15}. The study of the effects of numerical erosion will be left to future work.

The Interstellar Boundary Explorer (IBEX) was launched 19 October 2008 to study the heliosphere by observing energetic 
neutral atoms (ENAs) \citep{McComas09a}. ENAs provide an indirect method for studying the shape and thermodynamic properties of the heliosphere. IBEX is in a highly elliptical orbit around the Earth lasting roughly a week \citep{McComas11}, producing a global ENA map every six months for each energy band. There are two ENA cameras onboard IBEX: IBEX-Lo \citep{Fuselier09}, which measures ENAs from $\sim$10 eV to $\sim$2 keV, and IBEX-Hi \citep{Funsten09}, which measures ENAs from $\sim$300 eV to $\sim$6 keV. ENAs are created when an ion charge exchanges with a neutral atom, where the ion steals the electron from the neutral atom. Due to this charge exchange process, the energy of an ENA is dictated by its parent proton. Since different regions of the global heliosphere contain populations of ions with different energies, IBEX can probe these different regions by the energy signature of the ENAs. INCA was also able to observe ENAs while in orbit around Saturn. The primary objective of INCA was to observe ENAs originating from the plasma in Saturn's magnetosphere, but it was also able to image the heliosphere in directions away from Saturn \citep{Krimigis09, Dialynas13}. INCA imaged heliospheric ENAs in the $\sim$5 to $\sim$55 keV energy range, complementing observations of IBEX while also allowing a deeper look down the heliotail at energies $>30$ keV, where the charge exchange rate drops.

Shortly after its launch, IBEX observed a ribbon-like structure in its ENA observations \citep{McComas09b}. 
This ribbon, known as the IBEX ribbon, is superposed on top of a globally distributed flux (GDF) of ENAs generated by 
processes within the inner heliosheath (IHS) and the LISM. \cite{Schwadron11} first attempted to separate the IBEX ribbon from the GDF by applying a mask over the region immediately surrounding the ribbon and interpolating with respect to the background flux. From this, ENA maps of both the GDF and the IBEX ribbon could be made separately. Due to this technique, ENAs originating in the IHS can be studied using the GDF maps. This technique was used again by \cite{Schwadron14} to study the GDF over the five years of IBEX data, which revealed how solar wind affected ENAs in the IHS at different energies. From a theoretical perspective, this separation of the GDF from the IBEX ribbon was done by \cite{Heerikhuisen14}, who investigated the ENA contributions from both the IHS and the interstellar plasma disturbed by the heliosphere to better understand the ribbon creation mechanism and to simulate the ribbon. \cite{Zirnstein15} also studied both the GDF and the IBEX ribbon from a theoretical perspective, relating it to the effects of solar cycle variations and focusing primarily on the effects in the nose direction.

The first detailed analysis of IBEX tail measurements was published by \cite{McComas13}. The ENA maps showed two high latitude 
lobes in IBEX-Hi measurements, with an excess of flux from $\sim$2 keV 
to $\sim$6 keV. Additionally, at these energies a deficit of ENAs were also observed in the lower latitude heliotail
\citep{McComas13, Schwadron14}. Unlike the ribbon, it is believed that the lobes seen in the 
heliotail originate from the IHS at these energies. \cite{McComas13} proposed that the observed lobe structures 
could be attributed to the fast and slow solar wind in the heliosphere. During solar minimum, the fast solar wind is 
deflected towards high latitudes while the slow solar wind exists at lower latitudes. It was suggested that the high 
latitude lobes exhibit an excess of ENA flux at higher energies since they originate from the fast solar wind. Likewise, 
the deficit of ENAs in the lower latitude heliotail was attributed to the presence of slow solar wind.

This hypothesis of fast and slow solar wind being responsible for the lobes in the heliotail was investigated by \cite{Zirnstein16a} using the first five years of IBEX observations. \cite{Zirnstein16a} used a simple flow model of the heliosphere to simulate the deficit of ENAs in the lower latitude heliotail observed by IBEX at higher energies. While the presence of the fast and slow solar wind was shown to be a contributor to the lobes, it has not yet been shown that the fast/slow solar wind is entirely responsible for features in the tail. They found that the deficit was a result of the asymmetry in the solar wind. This study was followed by \cite{Zirnstein17} who created ENA maps of the heliotail after incorporating solar cycle dependence and extinction of pick-up ions (PUIs) in the IHS. Using a 3D MHD solution from \cite{Heerikhuisen13}, they were able to model the lobe structures seen in IBEX GDF measurements, but their model underpredicted the ENA flux. Here we intend to investigate how the ``croissant" heliosphere affects the GDF for a uniform solar wind. We first study the effect of ENAs from the confinement of the heliosheath flows by the solar wind magnetic field, a feature of the heliosphere that exists regardless of whether there is a comet-like tail or a split tail such as the ``croissant" heliosphere \citep{Michael18}. \cite{Izmodenov15} show similar confinement of the heliosheath plasma by the solar magnetic field, as evidenced by Fig. 6b in their work, which shows a peak in the mass flux around the azimuthal solar magnetic field; however, no synthetic ENA maps were created from this model. In the future, we will incorporate solar cycle time dependent effects as in \cite{Michael15}. We center maps on the nose as well as the tail to investigate both directions.

\cite{Opher15} used a line of sight integration of the ion pressure multiplied by the neutral density to create a 
proxy for ENA integration. In this proxy map, it is shown that for high energy ENAs, two regions of strong emission should
manifest themselves in the north and south. These regions of excess ENA emission show similar features to the ENA maps
from \cite{McComas13}. 

Due to the cooling length at the energies of IBEX, we are only able to probe until certain distances. The cooling length is the distance out to which 1/e of the local ions have survived charge exchange along a streamline. At the energies between 1.7 keV - 6 keV, within the range of IBEX, the cooling length is around 100 AU (Fig. \ref{Stream}b) so we cannot observe the tail at distances much beyond this unless there is an additional mechanism that either scatters high energy ions into lower energies or drives low energy ions to higher energy. Turbulence within the heliospheric jets might act as an energy driver, but will not be included in the ENA analysis presented here. On the other hand, above 10 keV, the charge exchange cross-section drops exponentially \citep{Lindsay05}, increasing the cooling length. Therefore, at energies $>$50 keV such as those measured by INCA, the cooling length (Fig. \ref{Stream}b) is expected to increase to distances greater than 200 AU. While INCA may have been able to probe deeper down the tail, IBEX is able to make a more complete global ENA map. ENA maps from INCA required the removal of Saturn's magnetospheric contribution and at times the solar direction, whereas the maps of IBEX are corrected for Earth's magnetospheric contribution as IBEX enters the magnetosheath in addition to other sources of noise. The effect of Saturn and the Sun on INCA measurements was greater than that which IBEX experiences, making IBEX observations better for comparing the structure of the heliosphere, though the lower energies prevent a thorough exploration of ENAs from the tail due to the short cooling lengths.

The purpose of this paper is to investigate the ENA maps produced by the ``croissant" heliosphere of \cite{Opher15}. We will explore the contribution of the heliospheric jets to ENA maps. In section 2, we present the model we use in calculating our simulated ENA maps. In section 3, we show our results. Finally in section 4 we discuss the implications of our results and how these results can drive future studies.

\section{Models}

\subsection{MHD Model}

We use a five-fluid model which is based on the 3D multi-fluid MHD code BATS-R-US \citep{Opher03, Opher09, Toth12}. The code uses one ion population with four neutral species and includes the magnetic field of the Sun as well as the LISM. Population 1 atoms are neutral atoms which undergo charge exchange with the disturbed LISM behind the bow shock. Population 2 atoms are those which are created within the IHS, in the region of compressed solar wind plasma downstream of the termination shock. Population 3 atoms are those which are generated within the supersonic solar wind. Finally, Population 4 atoms are in the pristine LISM and are of interstellar origin. 

The outer boundary of the domain is placed at $x={\pm}1500$ AU, $y={\pm}1500$ AU, and $z={\pm}1500$ AU. For this paper, since we are primarily interested in exploring the effect of confinement, we use a uniform, spherically symmetric solar wind. At the inner boundary instituted at 30 AU, we use the conditions from \cite{Opher15} of $v_{SW}=417$ km s$^{-1}$, $n_{SW}=8.74 \times 10^{-3}$ cm$^{-3}$, and $T_{SW}=1.087 \times 10^{5}$ K for the solar wind. The model is based on the \cite{Opher15} model, which uses a unipolar solar magnetic field to mitigate the effects of artificial reconnection. This artificial reconnection is present in the heliospheric current sheet when using a dipolar magnetic field. The radial component of the solar magnetic field at the equator is $B_{SW}=7.17 \times 10^{-3}$ nT at 30 AU.

For the interstellar plasma, we assume $v_{LISM}=25.4$ km s$^{-1}$, 
$n_{LISM}=9.25 \times 10^{-2}$ cm$^{-3}$, and $T_{LISM}=7500$ K. For the 
interstellar neutrals, we assume $n_{H}=0.155$ cm$^{-3}$ and that the speed 
and temperature are the same as for the interstellar plasma. For the 
interstellar magnetic field, we use a magnitude of B=2.93 nT, and 
orientation of $34^{\circ}.62$ and $227^{\circ}.28$ in ecliptic latitude 
and longitude, respectively, from \cite{Zirnstein16b}.  While we use the 
parameters from \cite{Zirnstein16b}, the interstellar magnetic field 
model and conditions from \cite{Opher15} are able to better reproduce 
the heliospheric asymmetries, such as the termination shock, in a steady-
state solar wind simulation. We use the same grid as in \cite{Opher15}, 
with 3 AU resolution in the tail region of the IHS out to 1000 AU. This 
allows us to have high enough resolution to capture and resolve the lobes 
down the tail, from which we calculate the ENA signal.

\subsubsection{Ion Populations}

Since our MHD model has a single ion component, it combines the thermal ions and PUIs as a single component. Therefore, it is not able to directly simulate ENA generation, since most ENAs at the energies observed from IBEX are produced from charge exchange from PUIs. PUIs are created by charge exchange and exhibit different characteristics based on the region of the heliosphere in which they are created. Different works include different populations of ions and model them in different ways. \cite{Chalov03} modeled the PUI spectra in the IHS, focusing on their spatial evolution in the upwind direction. These spectra were then used to model ENA fluxes at 1 AU. \cite{Fahr00} hydrodynamically modeled PUIs via a 5-fluid model which included protons, hydrogen, PUIs, anomalous cosmic rays, and galactic cosmic rays. The inclusion of these separate ion components was shown to affect the termination shock and heliopause location. \cite{Scherer05} continued on the work from \cite{Fahr00} by producing a separate PUI component in their modeling, and modeling the cosmic ray populations by solving the \cite{Parker65} transport equation. \cite{Malama06} also modeled PUIs as a separate component, treating multiple PUI populations kinetically.

Within the IHS, there should be three main populations of ions: thermal solar wind ions, PUIs created in the supersonic solar wind, and PUIs created in the IHS. \cite{Zank10} first attempted to model ENAs via a multi-ion plasma by using a three-dimensional MHD-kinetic global model of the heliosphere to include solar wind ions, PUIs transmitted across the termination shock from the supersonic solar wind, and reflected PUIs which are initially reflected at the termination shock and are not transmitted until they gain enough energy from the motional electric field. The characteristics of each population were estimated at the termination shock in terms of their number density and temperature. These populations were modeled as individual Maxwellian distributions, and maintain constant number density and temperature ratios relative to the plasma within the IHS. \cite{Zirnstein14} extended the work from \cite{Zank10} by including an additional ``injected" PUI population. The injected PUIs are the PUIs created locally in the IHS. For every loss of a solar wind ion, transmitted PUI, or reflected PUI, the injected population would gain an ion at a much lower energy. The injected energy is based on the relative speed between the neutral particle flow and the plasma bulk flow, which is lower in the IHS, giving the injected PUIs an energy of roughly 0.1 keV. This injected PUI population would thus not be a significant contributor to the measured ENA flux at IBEX-Hi energies, and are therefore neglected from this work.

\subsection{Pick-up Ion Calculations}

\begin{table}[t]
\centering
\begin{tabular}{ccc}
\tableline
Ion Population & $n_{i}/n_{p}$ & $T_{i}/T_{p}$ \\
\tableline
Trasmitted PUIs & 0.208 & 2.21\\
Reflected PUIs & 0.040 & 10.58\\
Solar Wind ions & 0.752 & 0.16\\
\tableline
\end{tabular}
\caption{Density and temperature ratios in the nose direction. PUI densities and temperatures are varied around the termination shock using these values as a reference point (i.e. we normalize all the densities along the termination shock by values at the nose using Eq. \ref{eq:VasVar}, and the temperatures are varied to maintain conservation of thermal energy). The nose is located at $\theta=90^{\circ}$ and $\phi=180^{\circ}$.}
\label{tab:ion}
\end{table}

We use the ion populations from \cite{Malama06} to model PUIs in the heliosphere by taking density and temperature ratios at the termination shock for the different ion populations relative to the total plasma. \cite{Malama06} modeled the different ion components of the IHS and their properties in a Monte Carlo procedure. This type of approach is important because PUIs most likely do not have Maxwellian velocity distributions and a kinetic approach to modeling PUIs is appropriate. Within the IHS, they treated five different ion populations: solar wind ions, transmitted PUIs created in the supersonic solar wind via charge exchange with an interstellar neutral or an ENA, and PUIs locally created in the IHS via charge exchange with an interstellar neutral or an ENA. In this work, we extract density and temperature ratios from \cite{Malama06} in the nose direction immediately downstream of the termination shock for the solar wind ions and the two transmitted PUI populations originating in the supersonic solar wind via the WebPlotDigitizer software (https://automeris.io/WebPlotDigitizer). We use these extracted ratios to create three main ion populations within our own modeling: solar wind ions, transmitted PUIs, and reflected PUIs.

For the solar wind ions, we use the extracted density and temperature ratios from \cite{Malama06} for this population. To create transmitted PUIs within our model, we combine the two transmitted PUI populations from \cite{Malama06} by summing their densities. We then assume that 16\% of the transmitted PUIs are reflected at the termination shock, and from this we can calculate the reflected PUI density. To determine the temperature of these populations, we partition the total thermal energy from the MHD simulation among the different ions such that \citep{Zank10},

\begin{equation}
T_{p}=\left(\frac{n_{SW}}{n_{p}}\Gamma_{SW}+\frac{n_{tr}}{n_{p}}\Gamma_{tr}+\frac{n_{ref}}{n_{p}}\Gamma_{ref}\right)T_{p},
\label{eq:Zank}
\end{equation}

\noindent where $n_{p}$ and $T_{p}$ are the density and temperature of the plasma from the MHD simulation, $n_{i}$ is the density for the respective ion population, and $\Gamma_{i}$ is the temperature fraction for the respective ion population given by $\Gamma_{i}=T_{i}/T_{p}$, with $T_{i}$ being the temperature for the respective ion population. In the work of \cite{Zirnstein17}, they assume thermal pressure fractions for each population, which are held constant along the surface of the termination shock ($n_{i}/n_{p}\Gamma_{i}$=constant). \cite{Zirnstein17} assume thermal pressure fractions of 4\% for the solar wind ions, 50\% for the transmitted PUIs, and 46\% for the reflected PUIs. Using the work of \cite{Malama06} we find that the solar wind thermal pressure fraction is 12\% in our model. We assume the ratio of thermal pressure fractions between the transmitted and reflected PUIs is the same between our model and the work of \cite{Zirnstein17}, and we find thermal pressure fractions of 46\% and 42\% for the transmitted and reflected PUIs, respectively. We solve Eq. \ref{eq:Zank} to find the transmitted and reflected PUI temperatures.

As the ions originating from the supersonic solar wind move in the IHS, they are depleted due to charge exchange. Once they undergo charge exchange in the IHS, local PUIs are created in the IHS. In this work, we do not include the locally created PUIs. The density and temperature fractions at the termination shock can be found in Table \ref{tab:ion}. These temperature fractions are kept constant throughout the IHS, whereas the density fractions change as we include extinction for the ions, detailed in section 2.2.3.

While PUIs exhibit suprathermal tails and therefore the total proton distribution is more accurately described by a kappa distribution, \cite{Zank10} showed that a superposition of Maxwellian distributions for the different ion populations approximates the results of a kappa distribution at lower energies. \cite{Desai12} and \cite{Desai14} also showed that the superposition of three Maxwellian distributions for the separate ion populations can produce ENA fluxes which are comparable to IBEX observations. We thus apply Maxwellian distributions to model all of the included ion populations.

\subsubsection{Angular Dependence of PUIs}
To include the angular dependence of the PUI distribution in our model, we use the neutral solution from our MHD model. We add the four populations of neutrals from our global MHD model for each grid cell, and we compare the different neutral densities across the termination shock. While our MHD model uses a single ion plasma and therefore we cannot distinguish between solar wind ions and PUIs within the plasma, charge exchange still occurs within our MHD solution. The neutral solution will be directly affected by the charge exchange process between the plasma and interstellar neutrals within the supersonic solar wind, so we can use our neutral solution to mimic the angular variation of PUIs across the termination shock.

\begin{figure}[t!]
  \centering
  \includegraphics[scale=0.4]{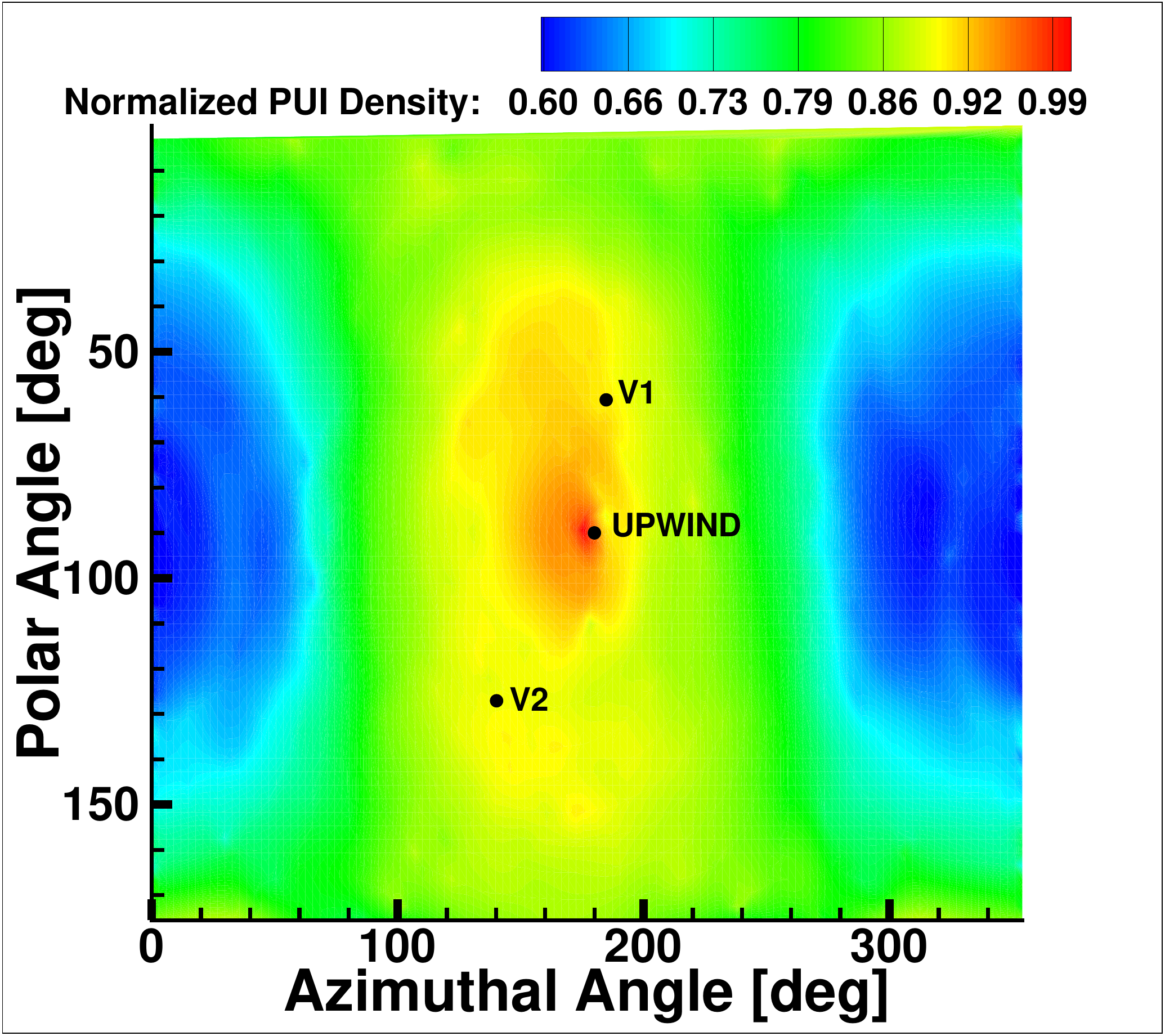}
  \caption{Angular variation of PUIs across the termination shock normalized to the nose calculated using Eq. \ref{eq:VasVar}. The PUI variation is based upon the multi-fluid neutral hydrogen solution from our MHD modeling, as well as the distance to the termination shock for each direction. The nose is located at $\theta=90^{\circ}$ and $\phi=180^{\circ}$.}
  \label{Var}
\end{figure}

We apply the variation in angle to the PUI populations in our model. We normalize the neutral hydrogen density across the termination shock by using the average neutral hydrogen density from within the supersonic solar wind in the nose direction as a reference point. The variation in the distance to the termination shock for a given direction will also cause the PUI density to vary. A longer distance to the termination shock means a solar wind ion will have a greater chance to charge exchange and form a PUI. We thus use the distance to the termination shock to vary the PUI density similar to the way we use the average neutral hydrogen density. Coupled with the density fractions from \cite{Malama06} and the total plasma density from our MHD solution, we can include an angular dependence of PUIs via

\begin{equation}
n_{PUI}(\textbf{r})=n_{i}(\textbf{r})\frac{n_{H,avg}(\theta,\phi)}{n_{H,avg}(\theta_{nose},\phi_{nose})}\frac{r_{TS}(\theta,\phi)}{r_{TS}(\theta_{nose},\phi_{nose})},
\label{eq:VasVar}
\end{equation}

\noindent where $\theta$ is the polar angle (latitude) and $\phi$ is the azimuthal angle (longitude). The polar angle increases from the northern pole towards the southern pole, while the azimuthal angle increases in the clockwise direction from the tail. The northern pole is at $\theta=0^{\circ}$ and $\phi=0^{\circ}$, and the nose is located at $\theta=90^{\circ}$ and $\phi=180^{\circ}$. Also, $r_{TS}$ is the distance to the termination shock for a given direction, $n_{i}$ is the PUI density calculated by using the total plasma density multiplied by the density fractions given by Table \ref{tab:ion}, and $n_{H,avg}$ is the average of the total neutral hydrogen density from the multi-fluid neutral solution from the inner boundary to the termination shock for a given direction. The angular dependence is shown in Fig. \ref{Var}. We also vary the temperature of the PUIs along the termination shock corresponding to the variation in PUI density such that the thermal energy, $n_{p}k_{b}T_{p}/(\gamma-1)$, is conserved. We find a much denser PUI population in the nose than in the tail as expected. The densities are normalized to the PUI density at the nose. With increasing angle from the nose, we see a decrease in the PUI density.

\begin{figure*}[t!]
\centering
\subfloat[]{\includegraphics[width=0.45\linewidth]{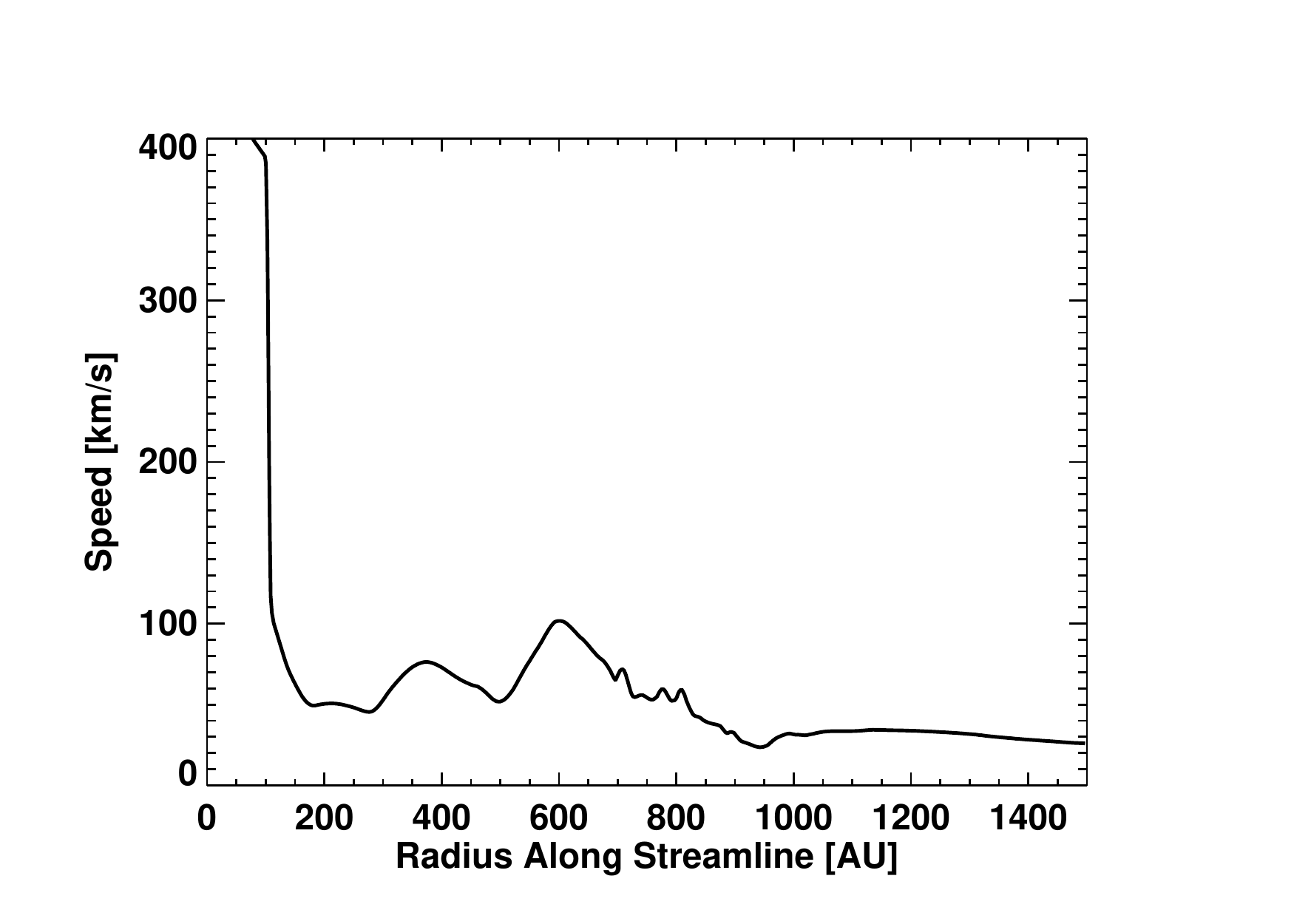}}  
\subfloat[]{\includegraphics[width=0.45\linewidth]{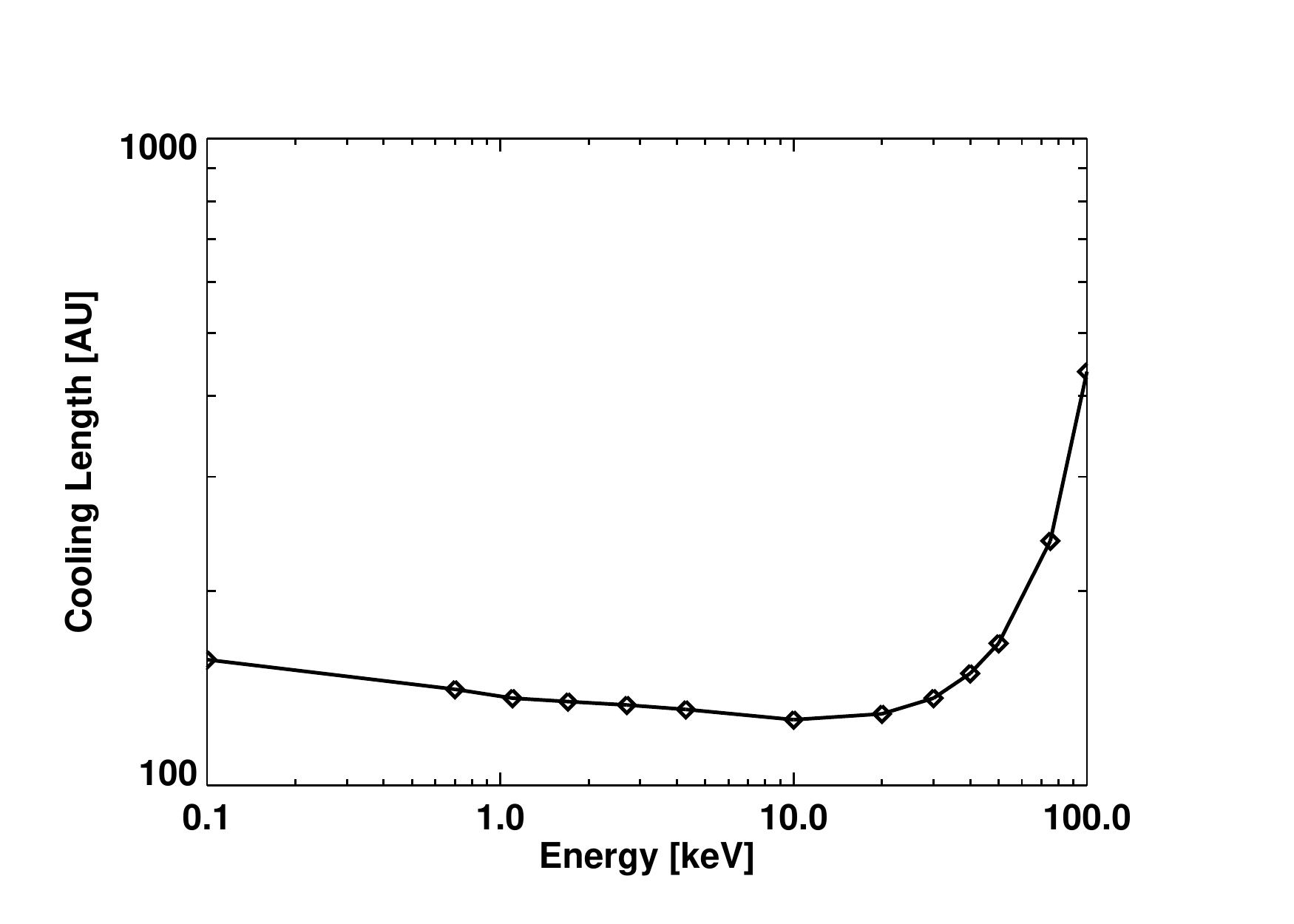}}  
\caption{Cooling Length and speed from a streamline extending down the tail. The streamline originates in the downwind direction at an ecliptic longitude of 77$^{\circ}$.84 and an ecliptic latitude of 22$^{\circ}$.21 at the termination shock. (a) The bulk plasma speed along the streamline. (b) The cooling length as a function of energy along the streamline.}
\label{Stream}
\end{figure*}

\subsubsection{Extinction}

The density fractions at the termination shock from \cite{Malama06} are used to calculate the ion fractions in the IHS for different radial vectors. Ions originating within the supersonic solar wind move beyond the termination shock and into the heliosheath, and are depleted due to charge exchange. Once they undergo charge exchange, local PUIs are created in the IHS. We take individual streamlines from the MHD solution and calculate the extinction of the PUIs starting from the termination shock using Eq. 7 of \cite{Zirnstein17},

\begin{equation}
n_{i}(\textbf{r})=\chi_{i}n_{p}(\textbf{r})e^{-\tau},
\label{eq:z17}
\end{equation}

\noindent $\mathbf{r}$ is the radial vector in space, $n_{i}$ is the ion density for population $i$, $\chi_{i}$ is the density fraction for population $i$, $n_{p}$ is the total plasma density, and $\tau$ is the extinction given by

\begin{equation}
\tau=\int_{r_{TS}}^{r} \frac{n_{H}(\textbf{r})\sigma(E)v_{i}(E)ds}{u_{p}(\textbf{r})},
\label{eq:ext}
\end{equation}

\noindent where $r_{TS}$ is the streamline distance to the termination shock, $n_{H}$ is the total neutral H density from the multi-fluid neutral solution, $\sigma$ is the charge exchange cross-section for a particular energy from \cite{Lindsay05}, $v_{i}$ is the speed of the parent proton which will yield an ENA of a particular energy, $u_{p}$ is the bulk plasma speed, and $ds$ is the path length over which we integrate the streamline. Figure \ref{Stream} shows both the bulk plasma speed and the cooling length at different energies along an example streamline extending into the heliotail. The streamline, picked as an example, originates in the downwind direction at an ecliptic longitude of 77$^{\circ}$.84 and an ecliptic latitude of 22$^{\circ}$.21 at the termination shock. As the streamline flows past the termination shock, it will change in longitude and latitude. This means that different cooling lengths of different energies along a given streamline will change in their longitude and latitude based on where the streamline is located in space.

The cooling length is influenced by the interstellar neutral density as well. Changing the interstellar neutral density at the outer boundary affects the depth out to which we can see down the tail. For a smaller neutral density, we can see further down the tail as the PUIs experience less charge exchange. Additionally, the energy of a particular ENA will influence how far down the tail we can see based on Eq. 2. In the IBEX-Hi energy range, the distance we can probe down the tail is limited ($\sim$100 AU), but at much higher energies ($\sim$100 keV) we are able to probe the tail much further up to distances of $\sim$400 AU. As the density of an ion population decreases, the locally created PUI density increases in the IHS, therefore making it a much more dense population further away from the termination shock. Since the locally created PUIs peak at $\sim0.1$ keV and we are focusing on energies greater than 1 keV, these locally created PUIs are negligible in our ENA calculations. Regardless, the locally created PUIs still have an effect on the calculated ENA intensities since they are directly caused by the extinction of the solar wind ions and transmitted PUIs. This extinction process depletes the transmitted ion populations at large distances.

\subsection{Calculating ENA Flux}

To calculate the ENA flux from the MHD model, we interpolate the Cartesian grid into a spherical grid. We interpolate the MHD plasma properties and neutral properties to a 2 AU $\times$ 3$^{\circ}$ $\times$ 3$^{\circ}$ spherical grid, with an outer boundary at 1500 AU. Since we are using a multi-fluid neutral approximation in our model, we smooth our solution where numerical effects occur. The ENA model is based on the model used in \cite{Prested08} and \cite{Opher13}, where we perform a flux integration along a radial line of sight. The equation for the ENA flux observed is given by

\begin{equation}
\begin{split}
J(E,\textbf{r}) =\int_{r_{observer}}^{\infty}\frac{2E}{m_{p}^{2}}f_{p}(n_{ion}(\textbf{r}'),T_{ion}(\textbf{r}'),v_{plasma}(\textbf{r}')) \\
n_{H}(\textbf{r}')\sigma(E)S(E)d\textbf{r}',
\end{split}
\label{eq:flux}
\end{equation}

\noindent where $m_{p}$ is the mass of a proton and $f_{p}$ is the phase space velocity distribution, which is treated as a Maxwellian for each modeled ion population. For the density and temperature of the given ion population, we use $n_{ion}$ and $T_{ion}$, respectively, where $n_{ion}$ is defined by Eq. \ref{eq:z17}, and $T_{ion}$ is a fraction of the local MHD temperature based on the thermal pressure fraction of the ion species. The velocity of the parent ion in the frame of the plasma, $v_{plasma}$, is given by $v_{plasma}=|\mathbf{v_{p}}-\mathbf{v_{i}}|$, with $\mathbf{v_{p}}$ and $\mathbf{v_{i}}$ being the velocities of the bulk plasma and the parent proton, respectively.

\begin{figure*}[t]
\centering
 \subfloat[]{\includegraphics[width=0.45\linewidth]{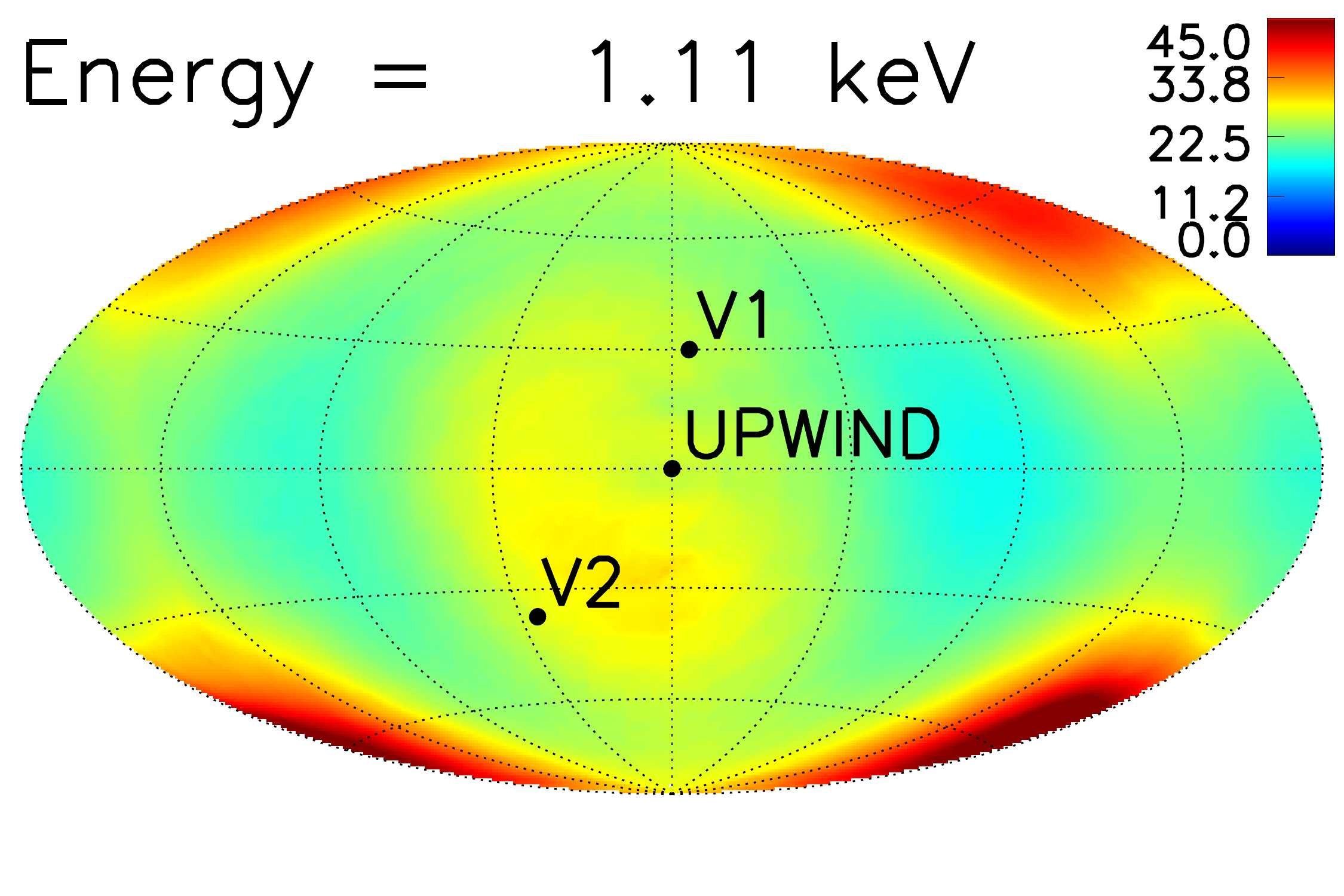}}  
 \subfloat[]{\includegraphics[width=0.45\linewidth]{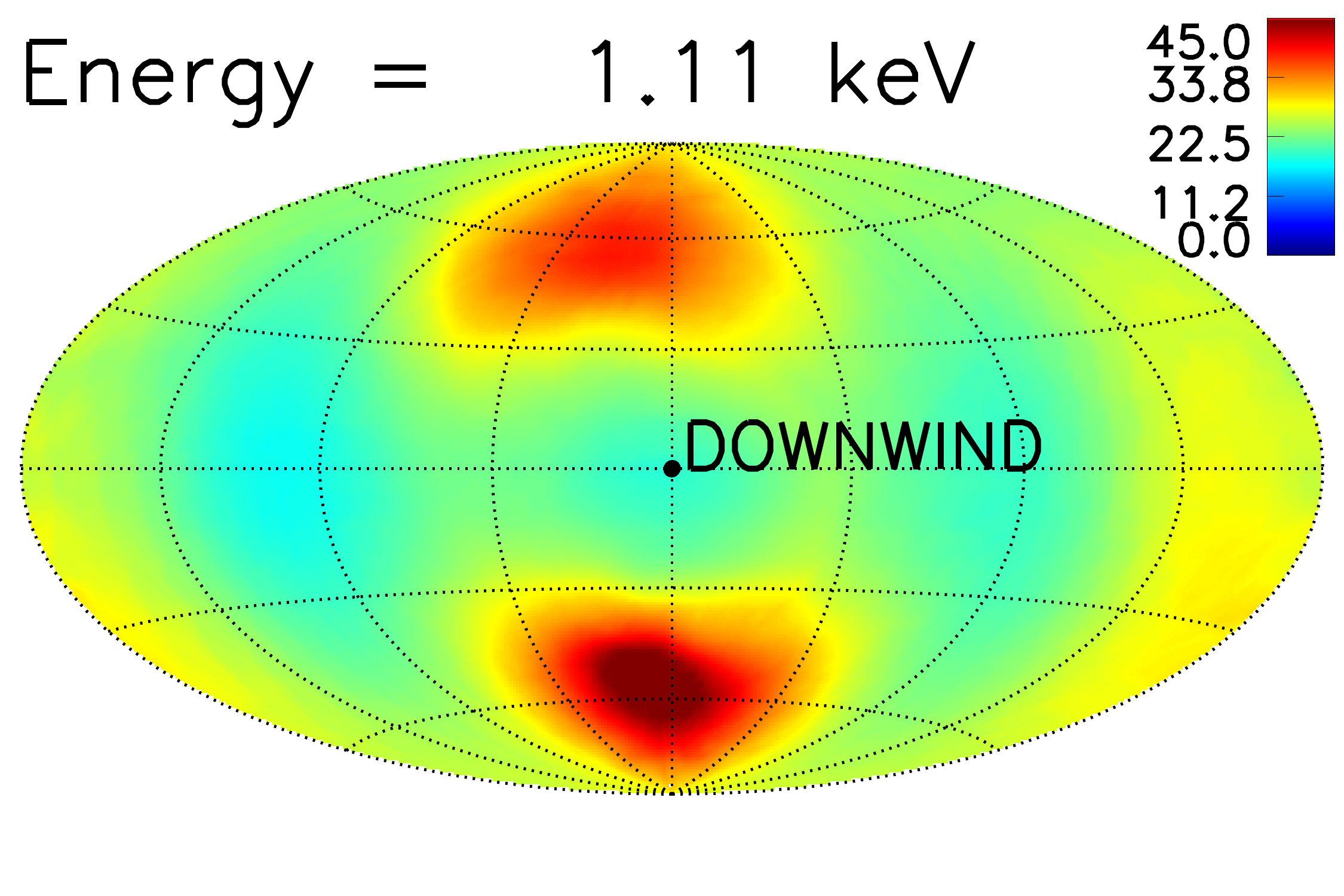}} 
 \\ 
 \subfloat[]{\includegraphics[width=0.45\linewidth]{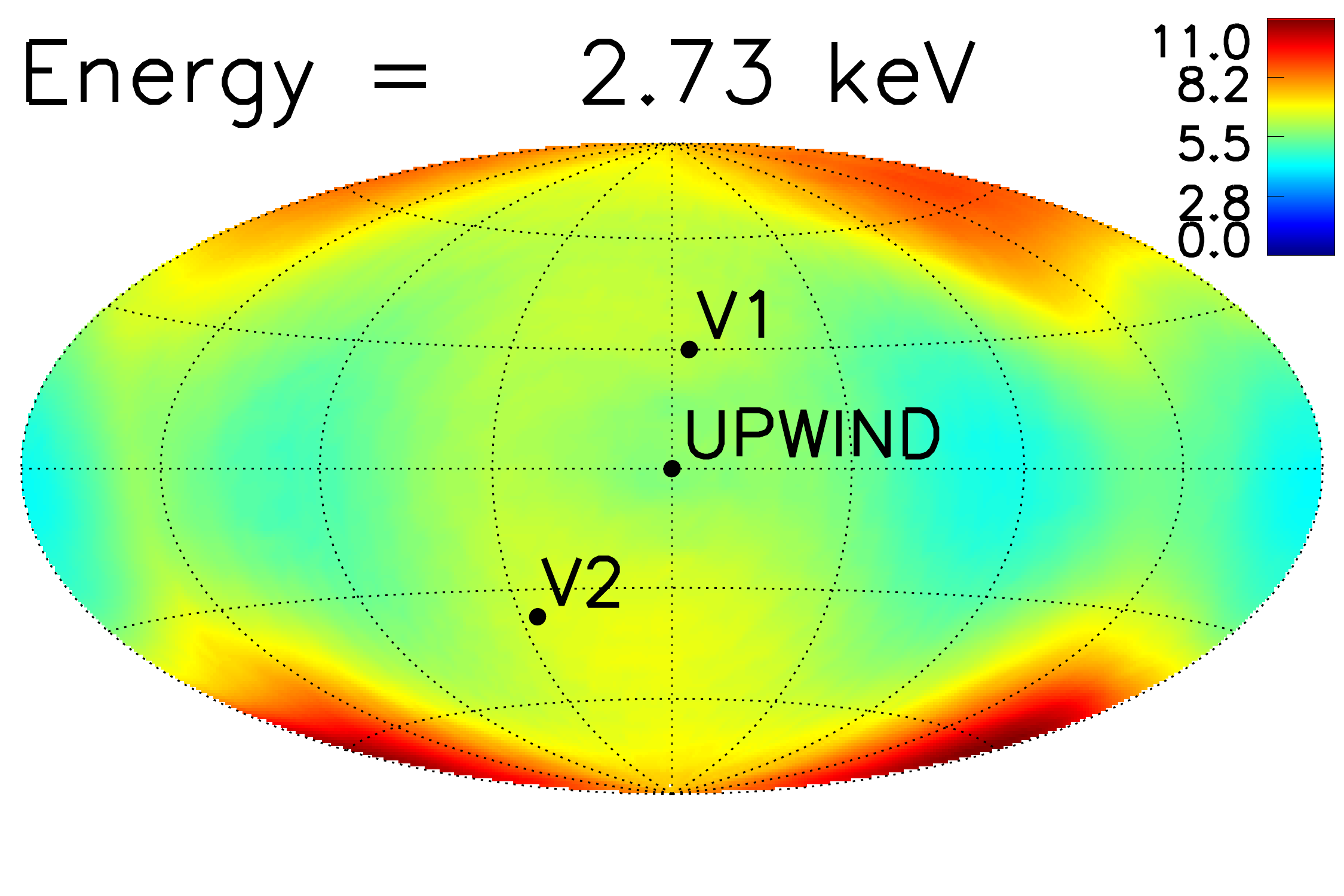}}   
 \subfloat[]{\includegraphics[width=0.45\linewidth]{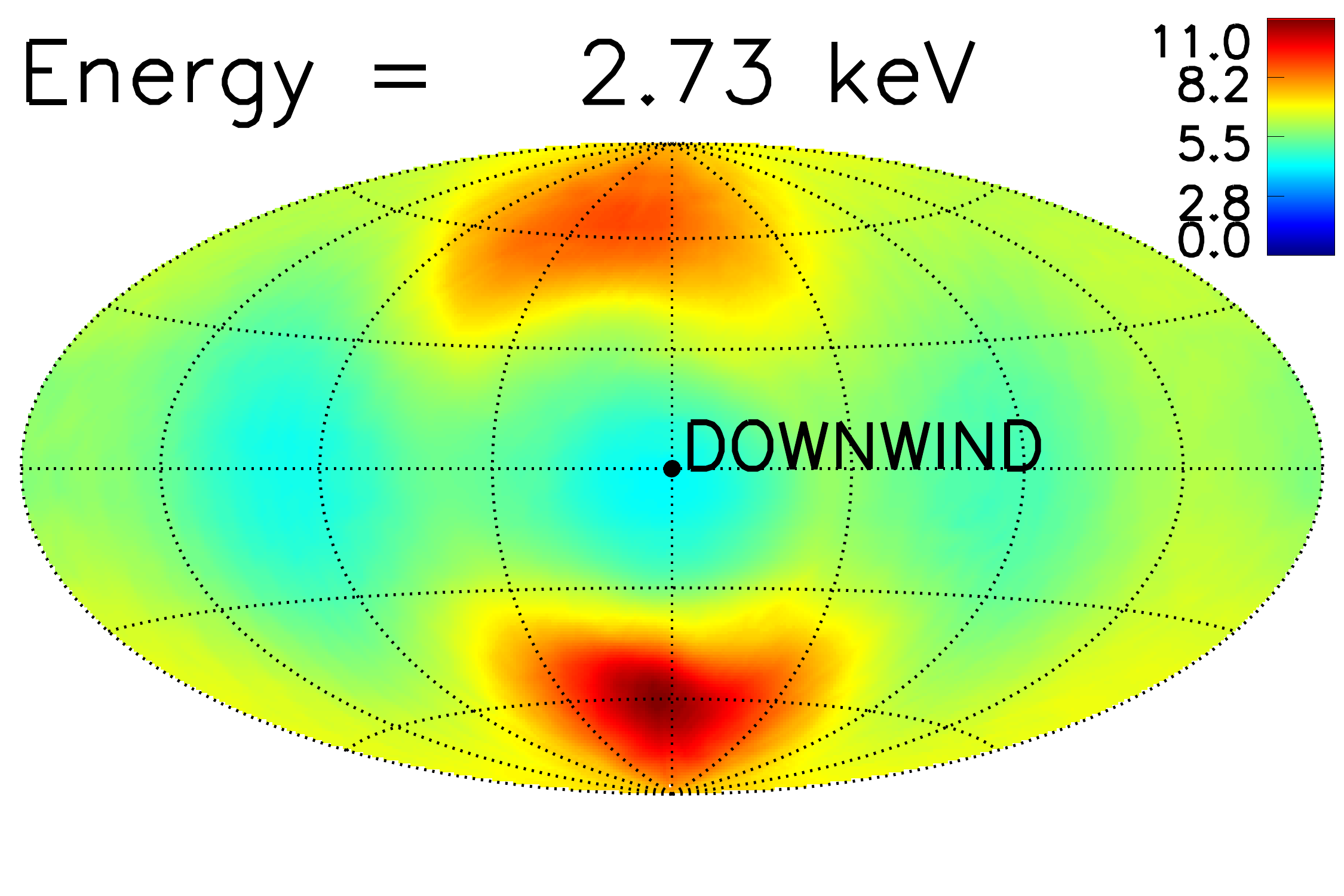}} 
 \\  
 \subfloat[]{\includegraphics[width=0.45\linewidth]{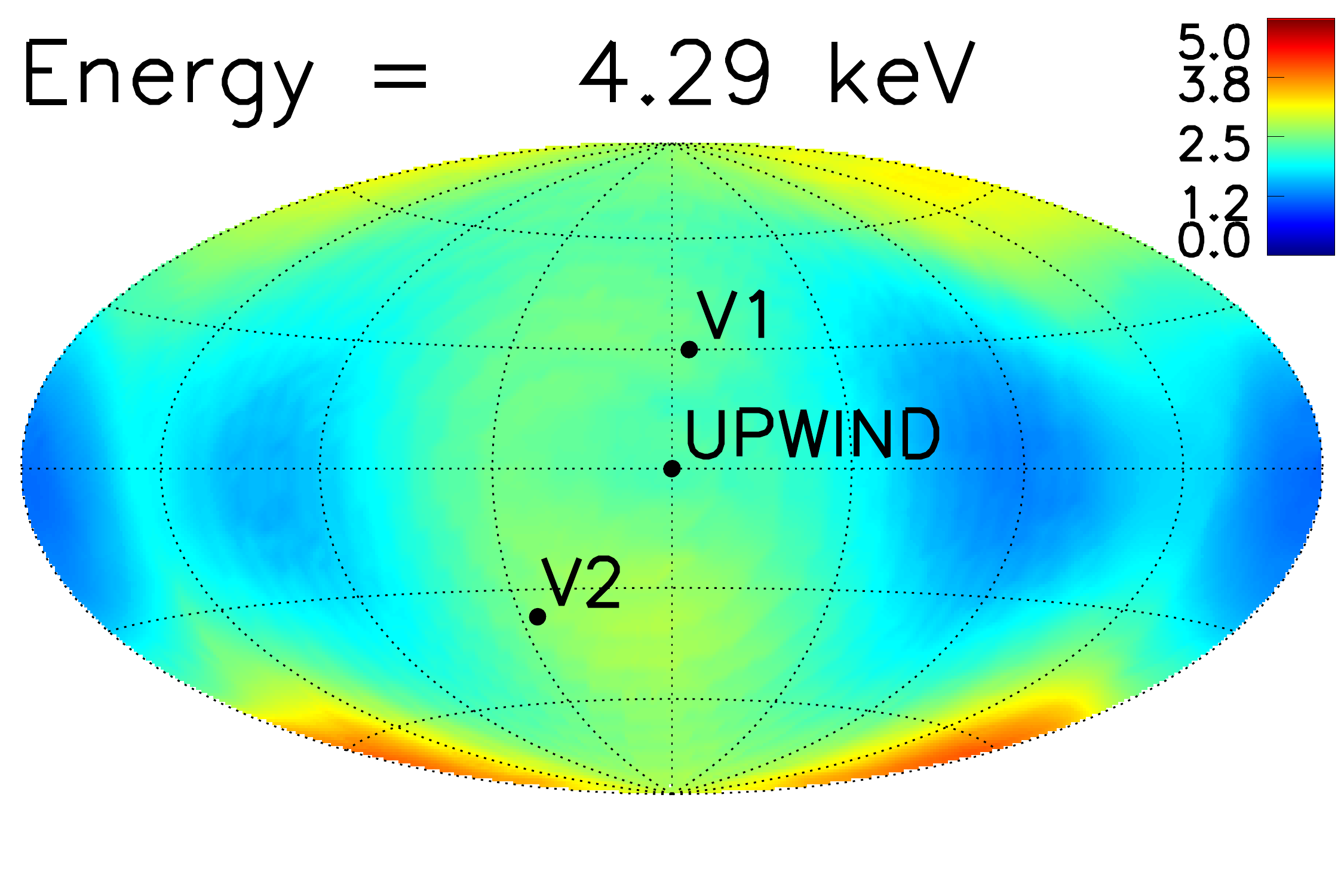}}   
 \subfloat[]{\includegraphics[width=0.45\linewidth]{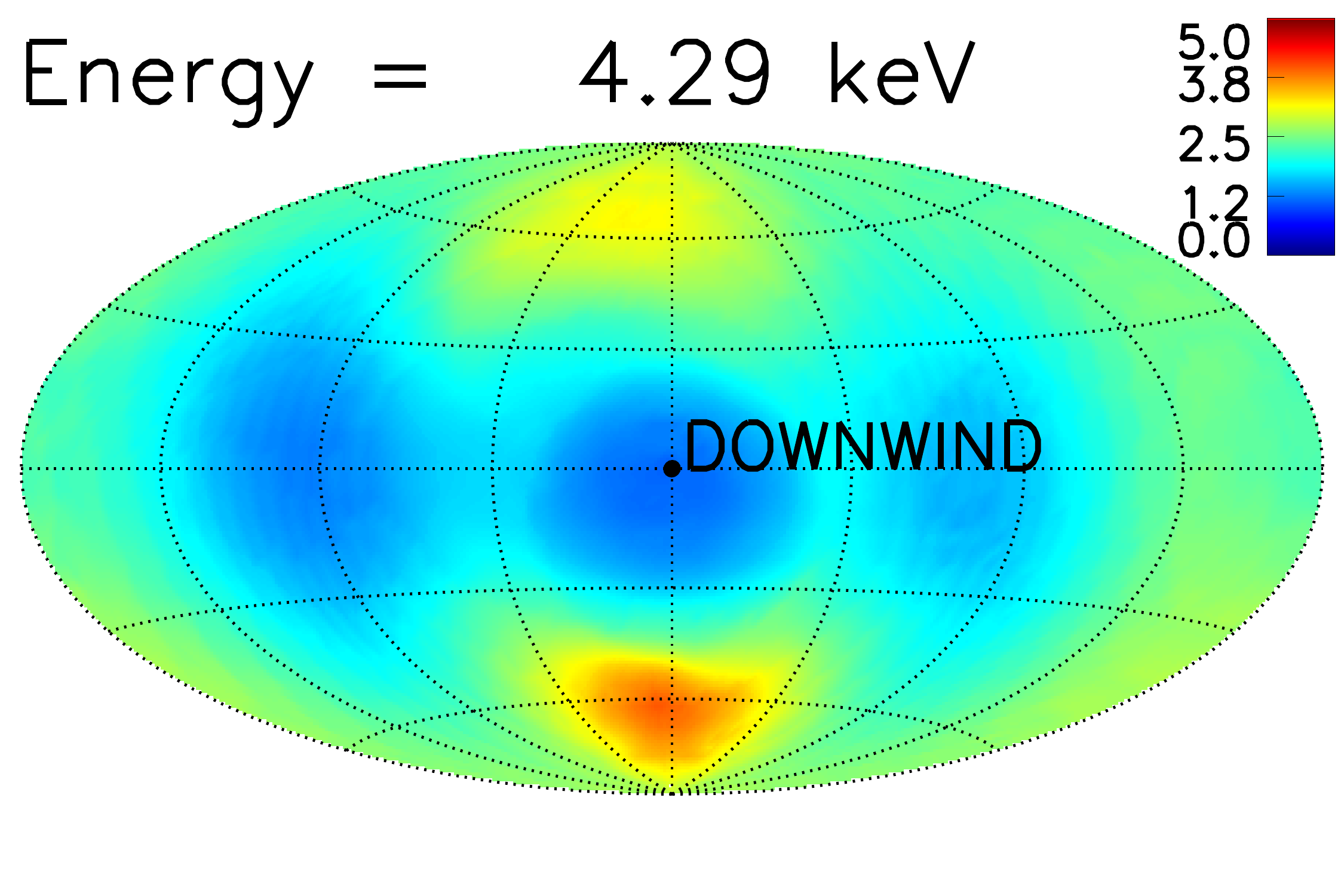}}   
\caption{Simulated maps of ENA flux with extinction in the IHS in units of (cm$^{2}$ s sr keV)$^{-1}$. The energy bands 
of the maps are centered on 1.11 keV (top), 2.73 keV (middle), and 4.29 keV (bottom). \textit{Left}: Nose-centered maps 
of ENA flux. \textit{Right}: Tail-centered 
maps of ENA flux.} 
\label{Maps}
\end{figure*} 

We also include survival probability, $S(E)$ in our ENA calculations. Survival probability is the probability that an ENA
will make it to an observer at a particular location \citep{Bzowski08}. This is calculated in a method similar to the extinction of ions 
used in the ENA creation process, except instead of calculating the charge exchange along a streamline, as we do for the ions, we calculate the charge exchange along a radial trajectory. Our survival probability for ENAs on radial trajectories back to Earth is given by

\begin{equation}
S(E)=\int_{r_{source}}^{r_{observer}} 
\frac{\sigma(v_{rel})v_{rel}n_{p}}{v_{ENA}}dr,
\label{eq:surv}
\end{equation}

\noindent where dr is the radial element over which we are integrating, $v_{ENA}$ is the speed of the ENA, and $v_{rel}$ is the relative velocity between the ENA and the bulk plasma given by \citep{Heerikhuisen06},

\begin{equation}
\begin{split}
v_{rel}= & v_{th,p}\left[\frac{e^{-\omega^{2}}}{\sqrt{\pi}}+\left(\omega+\frac{1}{2\omega}\right)erf(\omega)\right], \\
& \omega=\frac{1}{v_{th,p}}{\vert}\mathbf{v_{ENA}}-\mathbf{u_{p}}{\vert}.
\end{split}
\label{eq:rel}
\end{equation}

\noindent Here, $\mathbf{v_{ENA}}$ is the velocity of the ENA, $\mathbf{u_{p}}$ is the bulk averaged plasma velocity, and $v_{th,p}$ is the thermal speed of the plasma. The function to calculate relative velocities between the parent ions and the neutrals will change depending on whether a Kappa distribution or Maxwellian distribution is used. Equation \ref{eq:rel} is derived assuming the background plasma distribution has a Maxwellian distribution. In our MHD model we only include ionization via charge exchange. We place the observer at the termination shock, similar to IBEX ENA maps, which utilizes a survival probability correction out to 100 AU.

In our model, we define the IHS by $lnT_{p} > 12.8$, where $T_{p}$ is the plasma temperature. This captures the plasma in the IHS bound by the solar magnetic field. We do not model the ENA flux from beyond the IHS due to a lack of knowledge about the PUI characteristics in the ISM.   

\begin{figure*}[t]
\centering
  \subfloat[]{\includegraphics[width=0.45\linewidth]{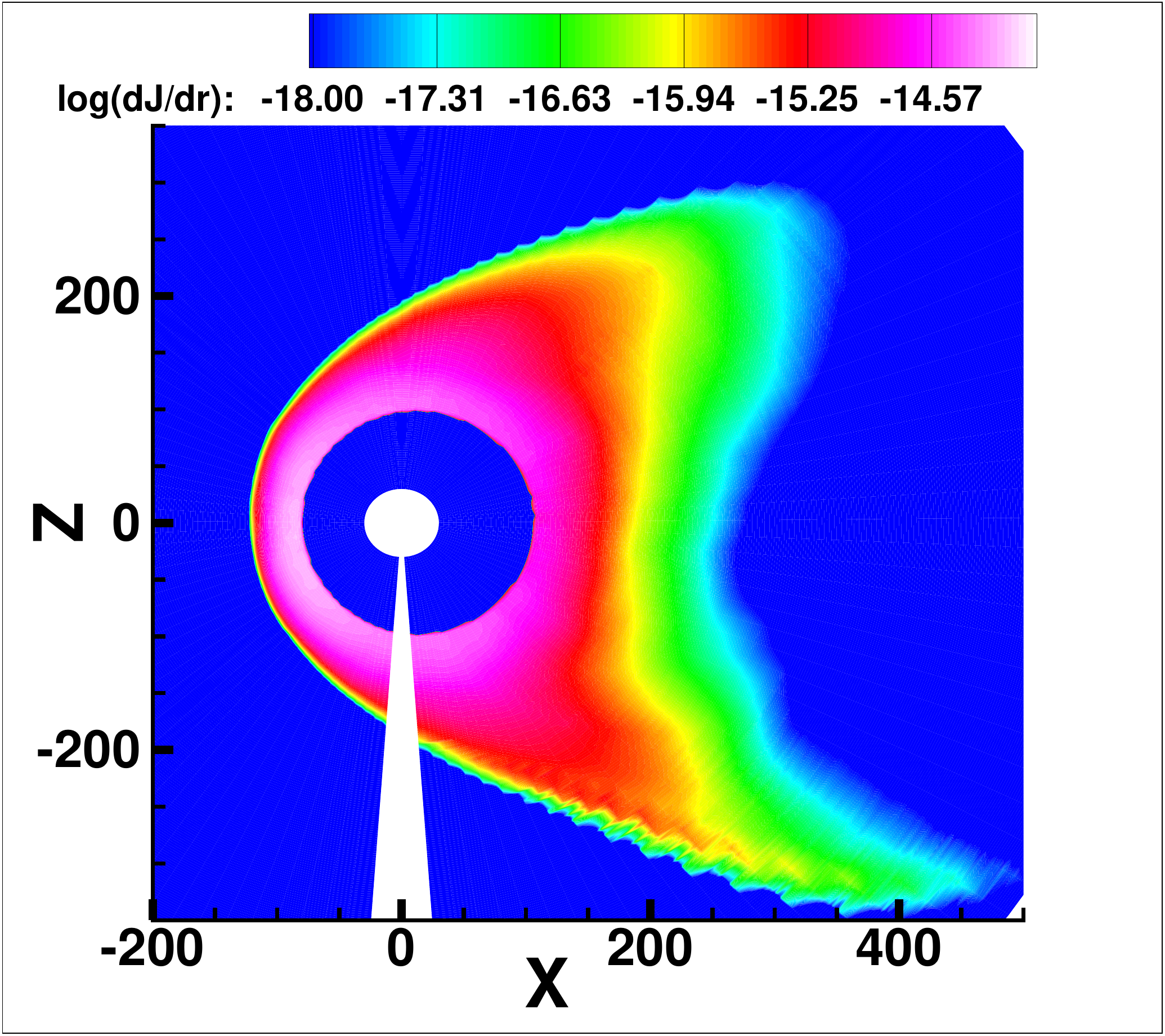}}
  \subfloat[]{\includegraphics[width=0.45\linewidth]{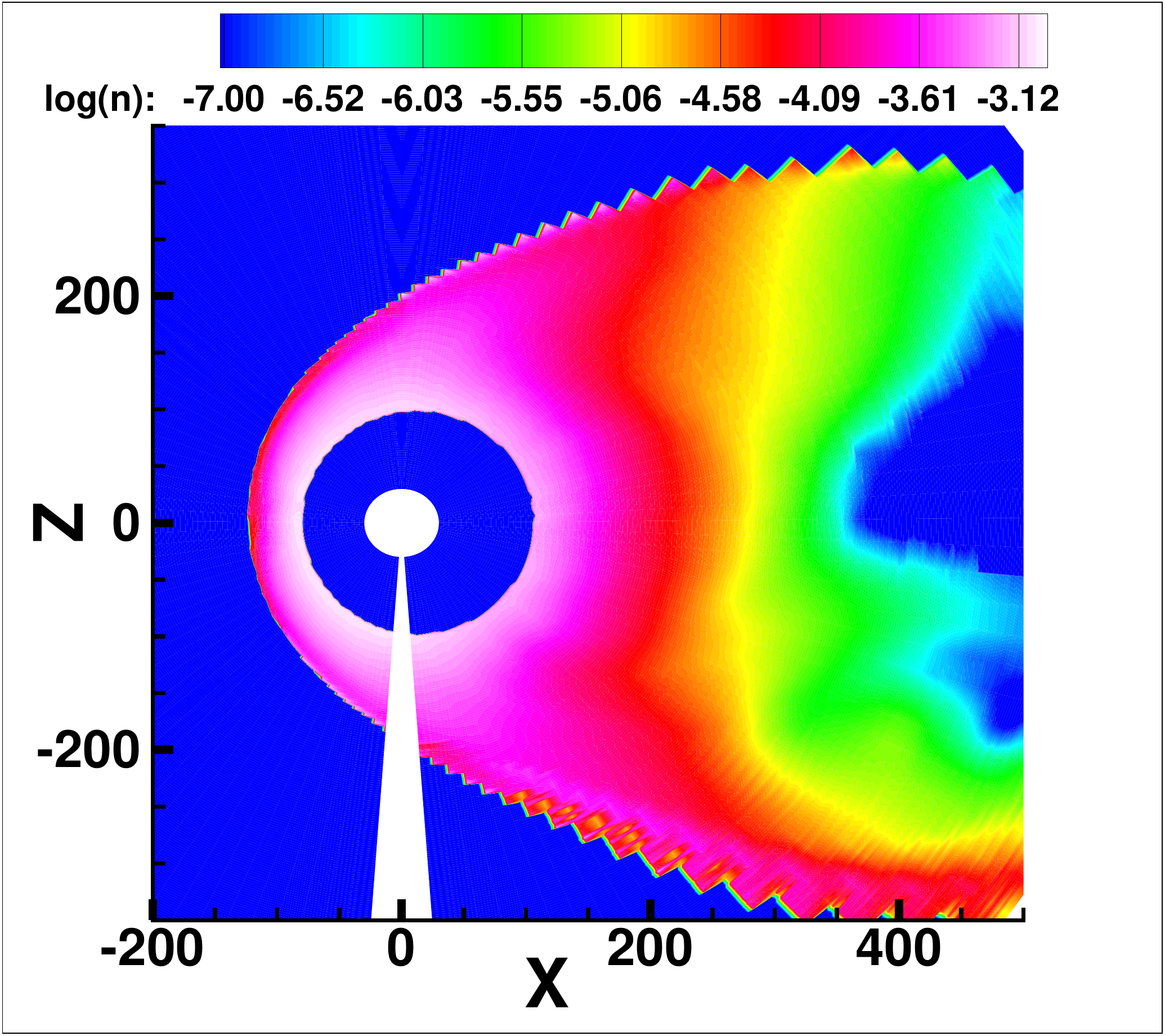}}  
  \\
  \subfloat[]{\includegraphics[width=0.45\linewidth]{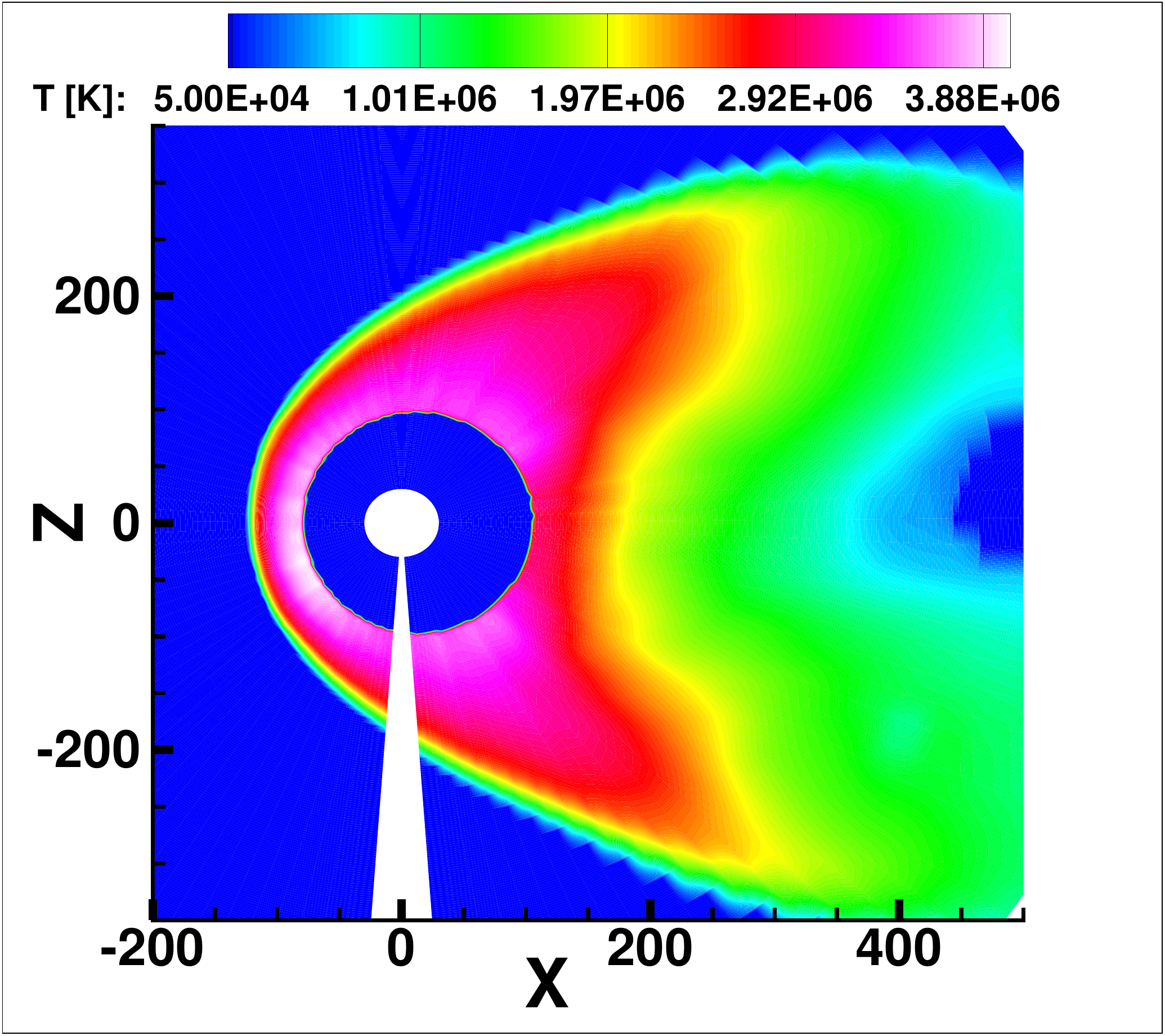}}   
  \subfloat[]{\includegraphics[width=0.45\linewidth]{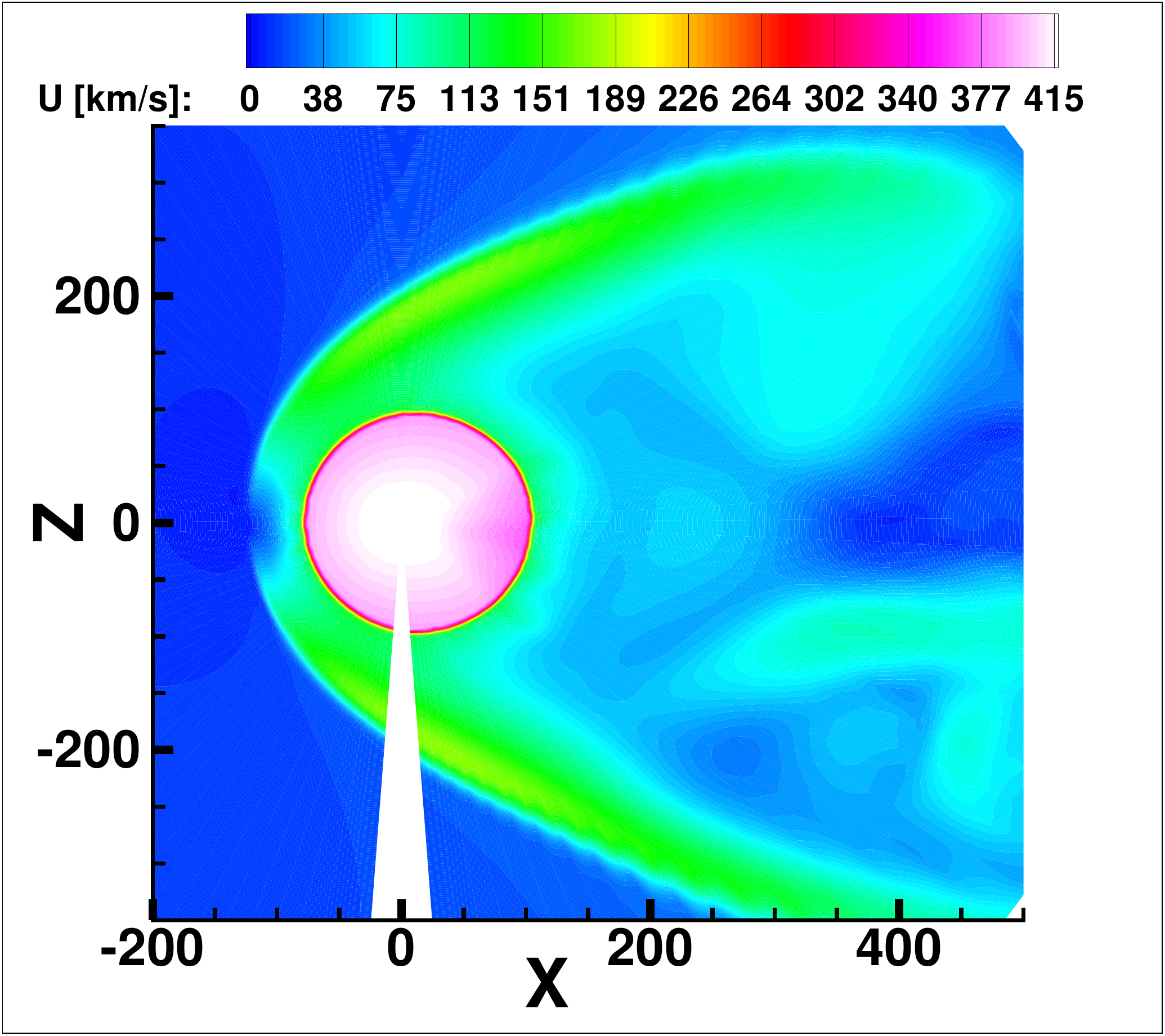}}   
 \caption{Meridional slices for the 4.29 keV energy band. The blank wedge is  an effect of the spherically interpolated 
 grid and does not affect our ENA maps.  (a) The log of the ENA flux with units of (cm$^{3}$ s sr keV)$^{-1}$ resulting from the transmitted PUI population originating from interstellar neutrals at each individual radius (dJ/dr). (b) The log of the density with  density units of cm$^{-3}$. 
 (c)  The temperature, and (d) the speed, U. All variables are shown for the transmitted PUIs.} 
\label{FluxComps}
\end{figure*}

\section{ENA Maps}

In Fig. \ref{Maps} we present the results of our simulated maps of ENA flux. There are three energy bands included, which are centered around 1.11 keV, 2.73 keV, and 4.29 keV to match the IBEX energy bands \citep{Funsten09}. For these maps, Eq. 5 is integrated to 600 AU in all directions. Our integration distance is limited to 600 AU because extinction removes nearly all of the parent protons at this distance in the IHS, and the ENA flux beyond 600 AU is negligible. We use the MHD model described in section 2.1 to simulate our ENA maps. The left panels of Fig. \ref{Maps} show the upwind-centered maps for the three different energy bands. We see an excess of flux towards the nose of the heliosphere as compared to the tail of the heliosphere. In the right panels of Fig. \ref{Maps}, we present downwind-centered maps, which show the presence of high latitude lobes towards the tail with an increase of the relative ENA flux.

\begin{figure*}[t]
\centering
  \subfloat[]{\includegraphics[width=0.45\linewidth]{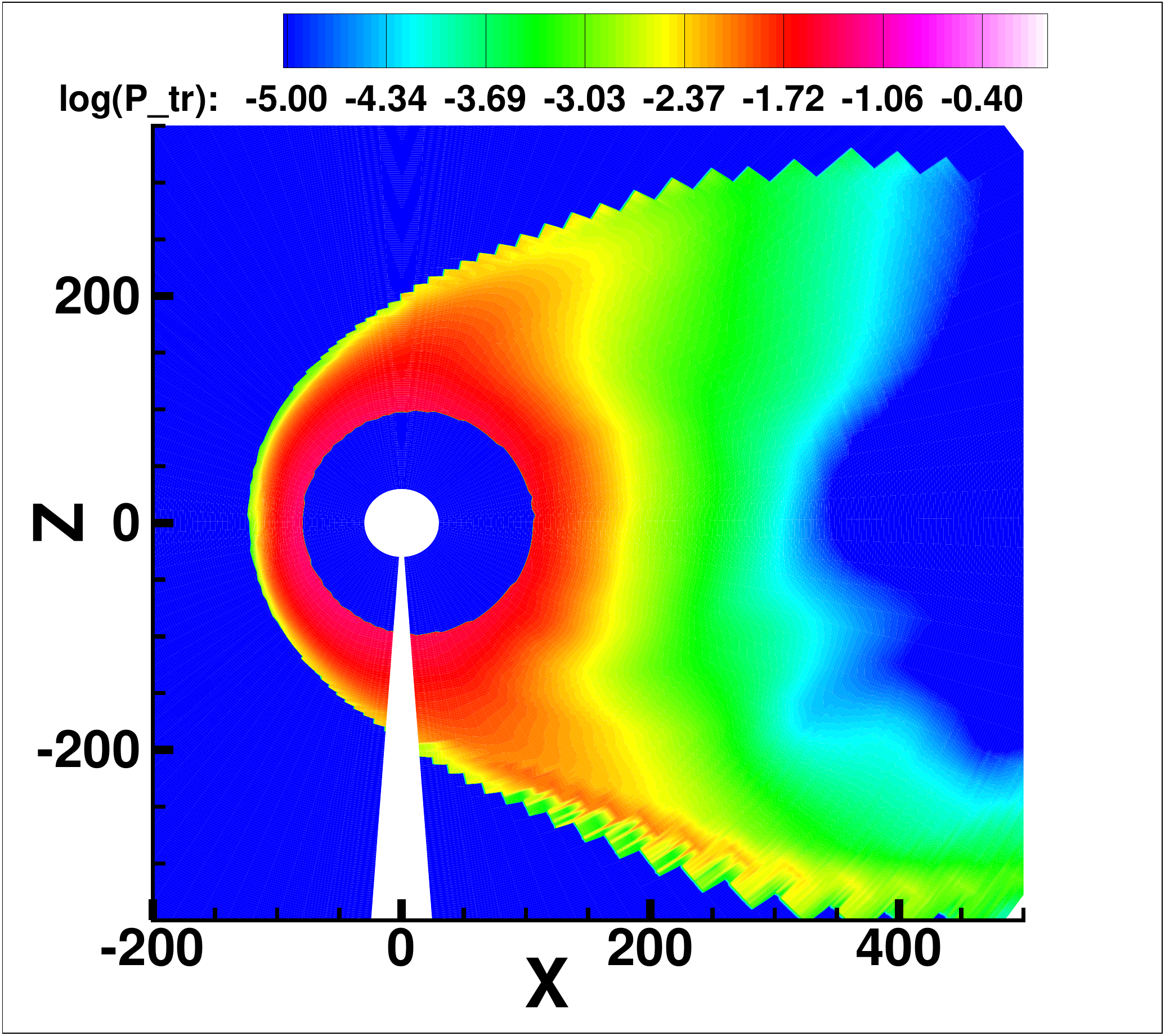}}  
  \subfloat[]{\includegraphics[width=0.45\linewidth]{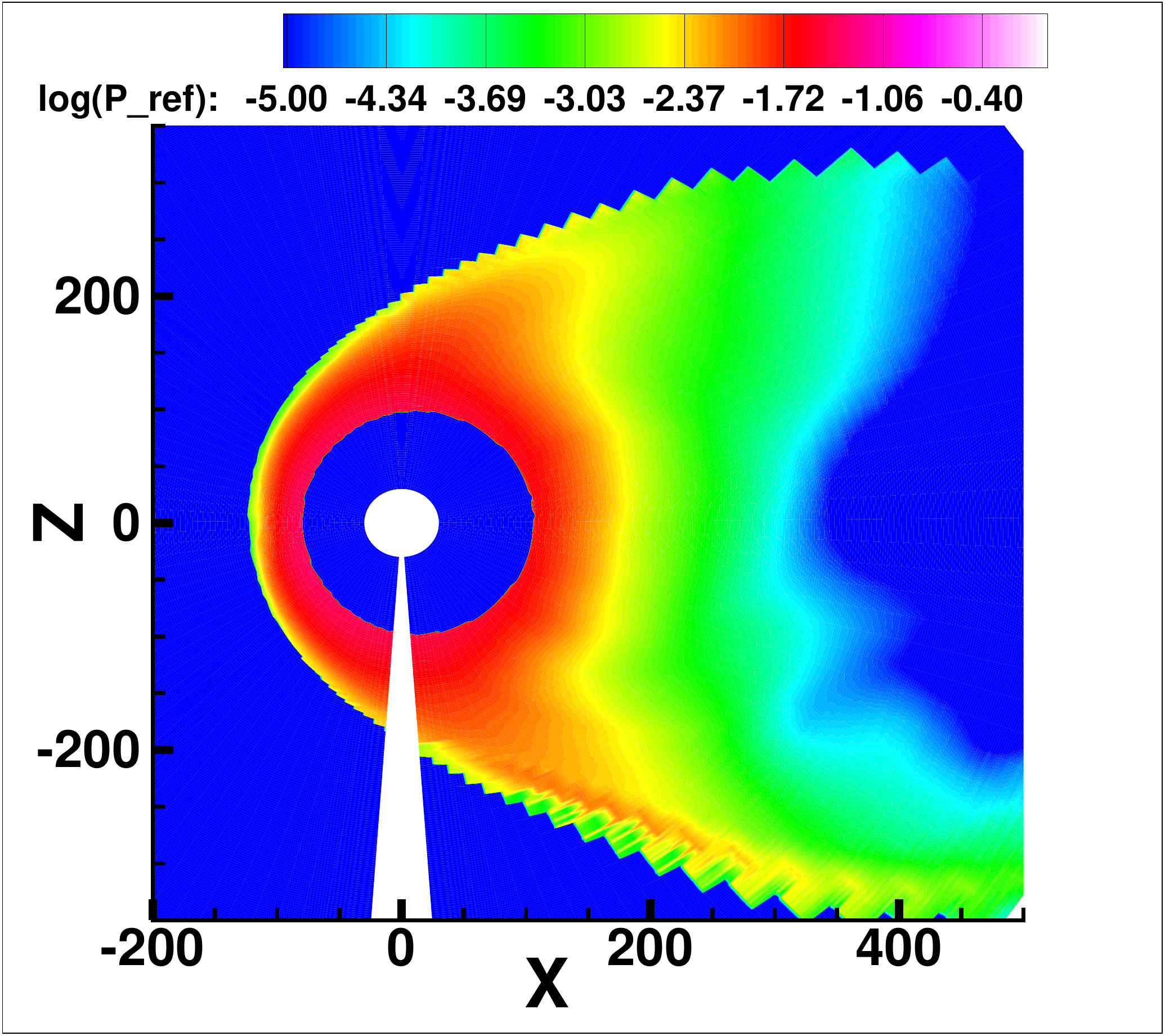}} 
  \\ 
  \subfloat[]{\includegraphics[width=0.45\linewidth]{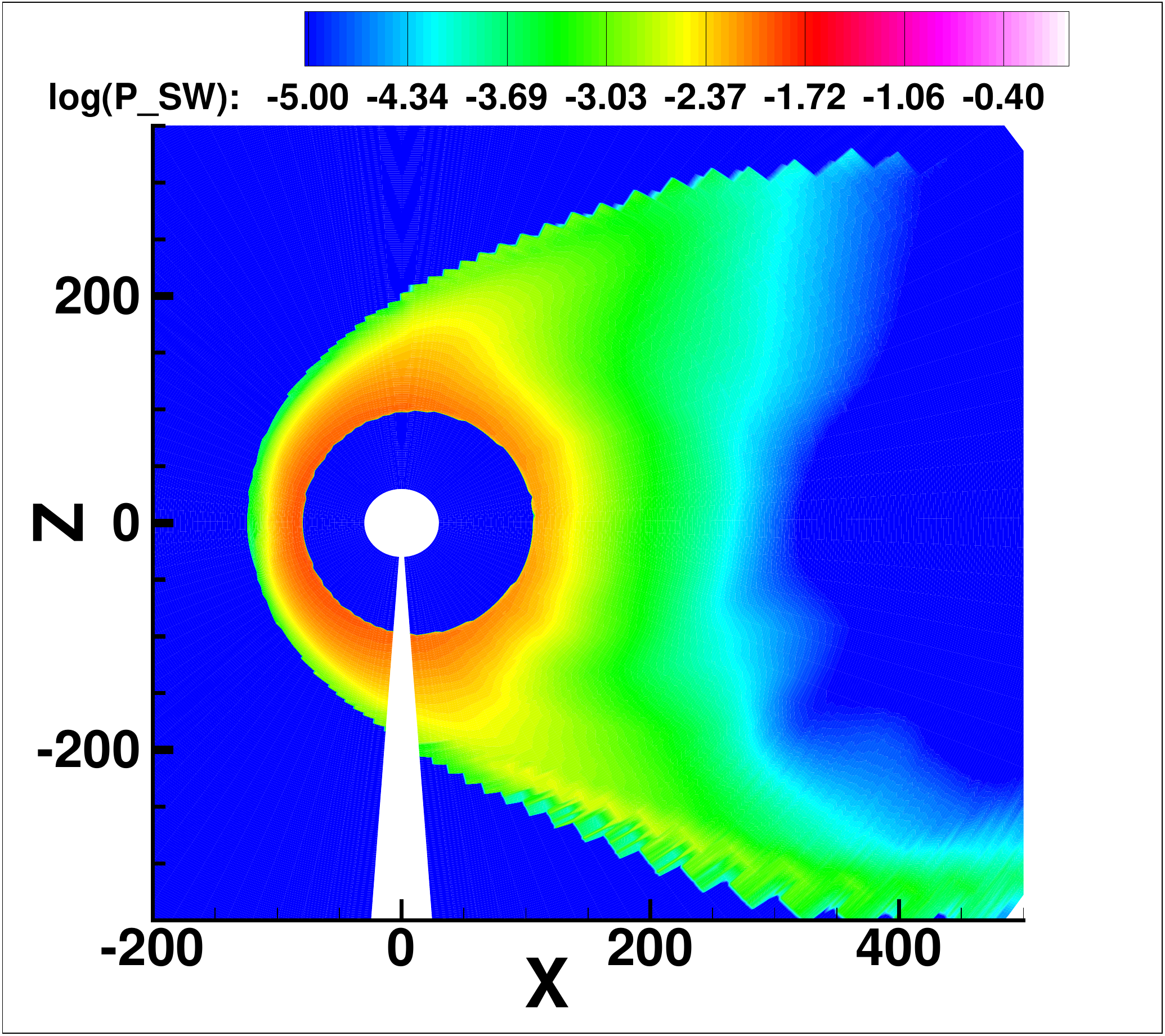}}   
  \subfloat[]{\includegraphics[width=0.45\linewidth]{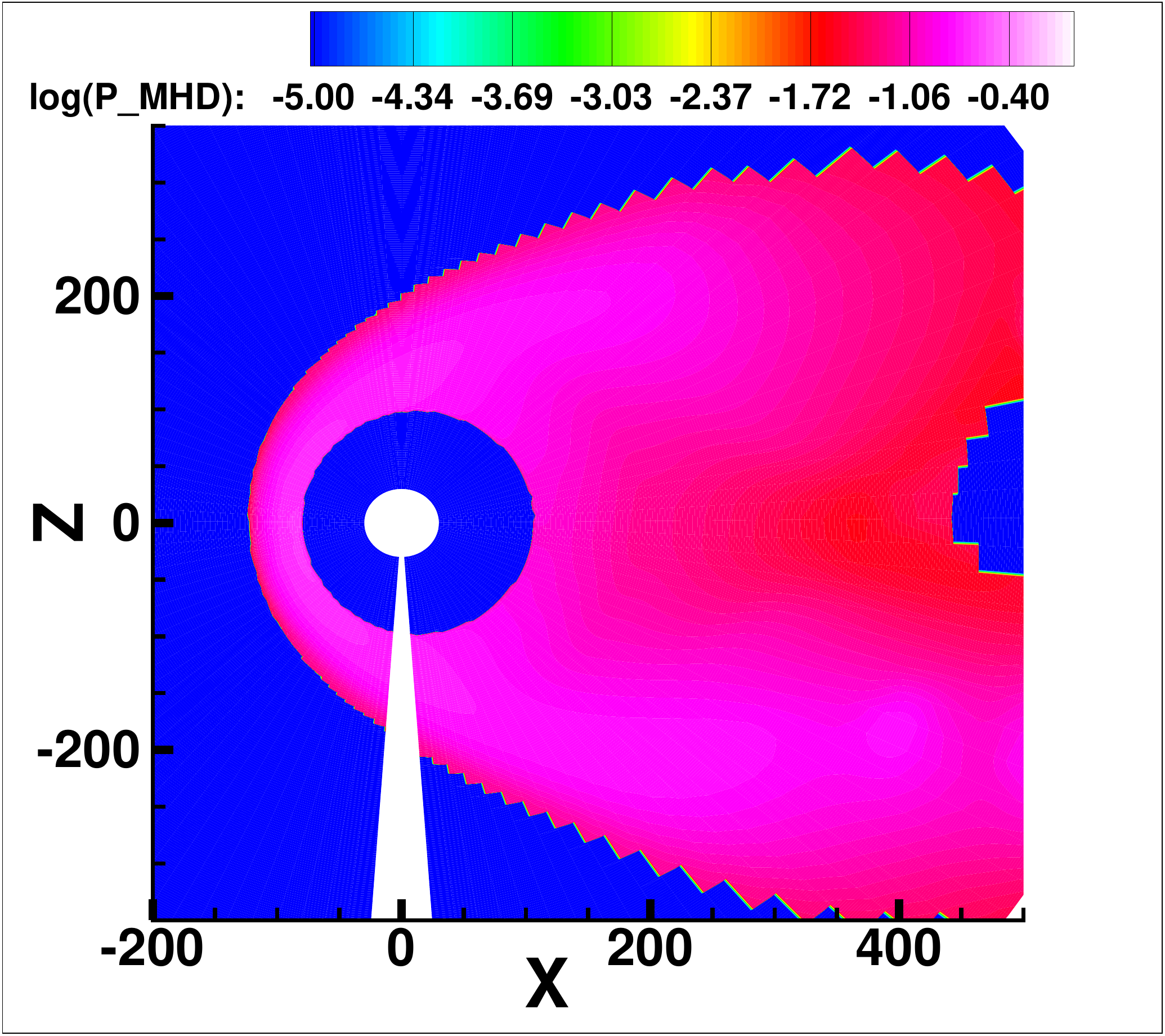}}   
 \caption{Meridional slices of pressure in units of log$_{10}$(pPa) at 
 the 4.29 keV energy band. The blank wedge is  an effect of the spherically interpolated grid.  (a) Pressure of transmitted PUIs,
 (b) pressure of reflected PUIs, (c) solar wind ion pressure, and (d) the plasma pressure from the MHD solution.} 
\label{Pressure}
\end{figure*}

At higher latitudes in the tail direction, we see two lobes of strong flux compared to other latitudes, which evolve across the energy bands in Fig. \ref{Maps}. This is due to the IHS material being funneled by the solar magnetic field. The funneling exists because the magnetic tension force of the solar magnetic field is able to resist the stretching caused by the heliosheath flows \citep{Opher15}. As further shown in \cite{Drake15}, the decrease in the plasma pressure between the termination shock and heliopause in the radial direction is controlled by the magnetic tension that funnels the heliosheath flows to the north and south. The funneling of the heliosheath plasma leads to an increase of the thermal pressure along the axis of the jet, yielding a high temperature at the poles. IBEX is able to see evidence of two high latitude lobes in ENA flux at the 2.73 keV and 4.29 keV energy bands \citep{McComas13}; however, these lobes are not seen in the lower energy maps with energy $<$2 keV. Our maps show the lobes persist in all energy bands. We will explore in a future work the effect of a time dependent solar wind on the confinement of plasma, and on the evolution of lobes in the ENA maps.

\cite{Opher09} and \cite{Izmodenov09} both show that the BISM direction can cause asymmetries in the heliosphere, such as the asymmetries between the Voyager crossings at the termination shock. The BISM can distort the heliosphere and orientation of the jets with respect to the north-south rotation axis of the Sun. This same asymmetry affects the jets in the ``Croissant" model, which can be seen in the differences between the ENA flux from the northern and southern lobes. 

We find that the high latitude lobes persist across all energies. We also note a deficiency of flux in the low latitude downwind directions relative to the high latitude lobes. The flux at the low latitudes between the lobes decreases relative to the lobes with increasing energy, and features a prominent structure spanning approximately 270$^{\circ}$ in longitude. In contrast, the IBEX maps centered on the tail show a flux enhancement of ENAs at lower latitudes in 0.71 keV, 1.11 keV, and 1.74 keV. This could be in part due to the latitudinal variation of the solar wind (fast and slow solar wind), which is not present in this work. Additionally, the presence of a time-dependent solar wind could also have a significant effect on the signal from high latitudes. 

\begin{figure*}[t]
\centering
  \subfloat[]{\includegraphics[width=0.45\linewidth]{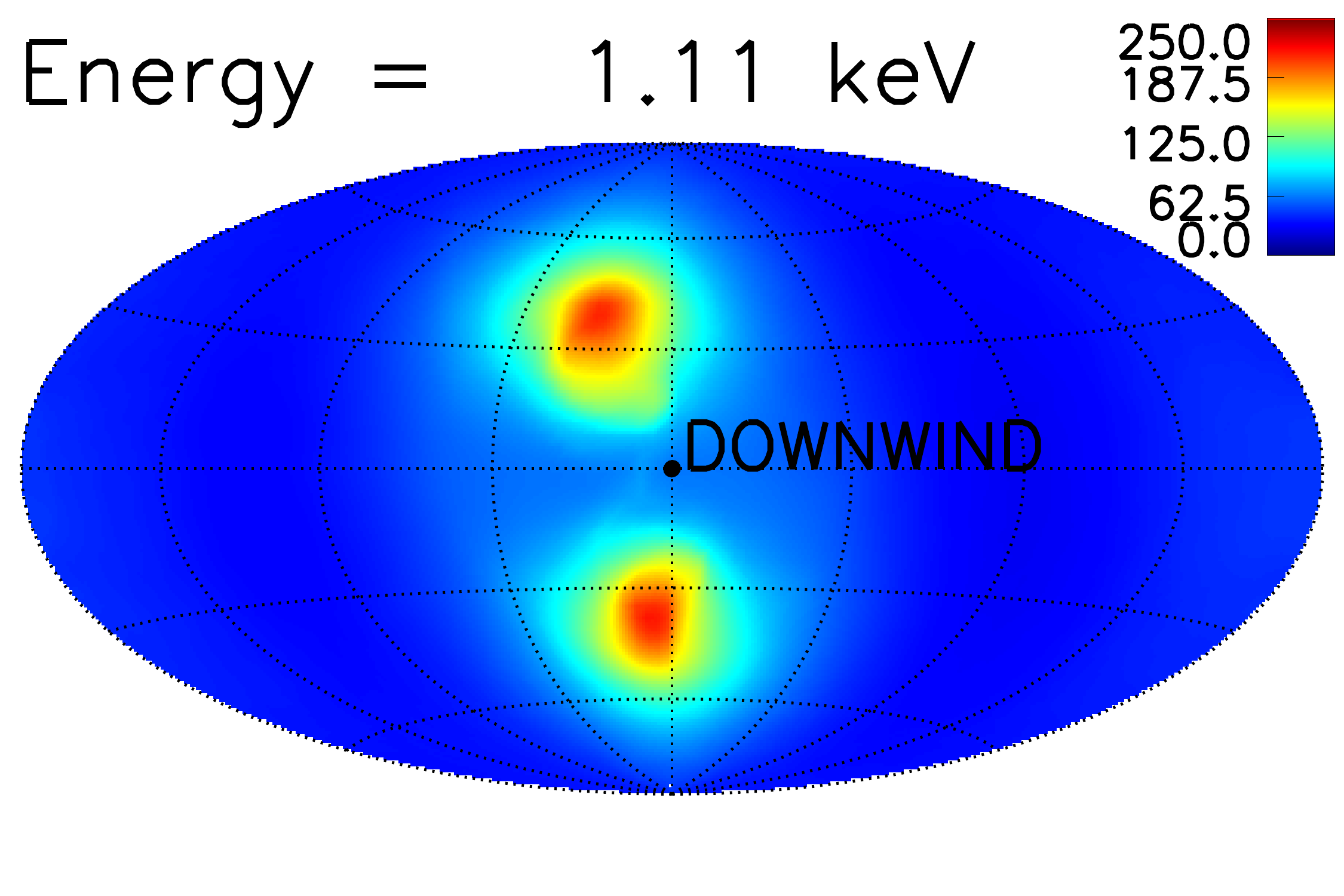}}  
  \subfloat[]{\includegraphics[width=0.45\linewidth]{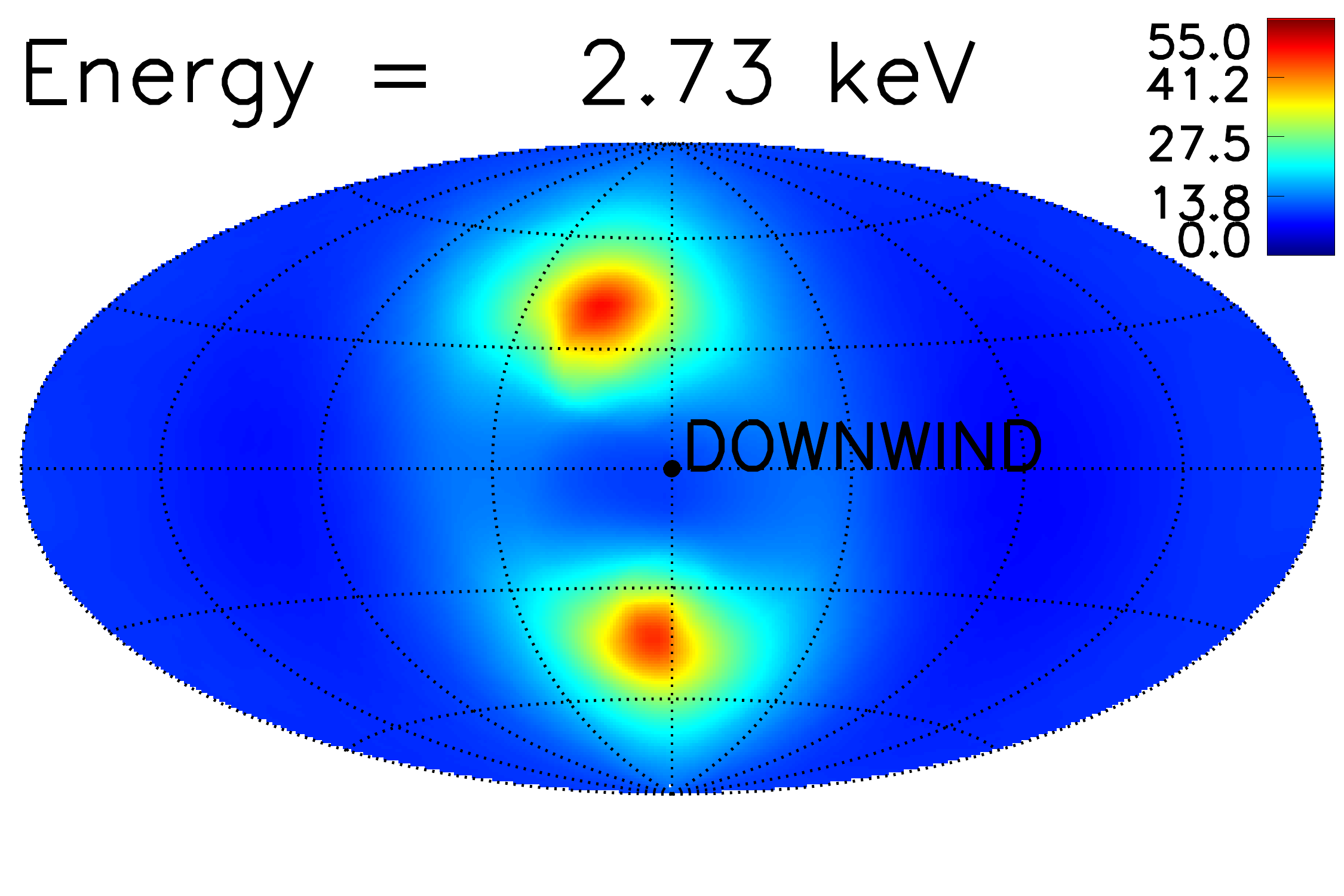}}  
  \\
  \subfloat[]{\includegraphics[width=0.45\linewidth]{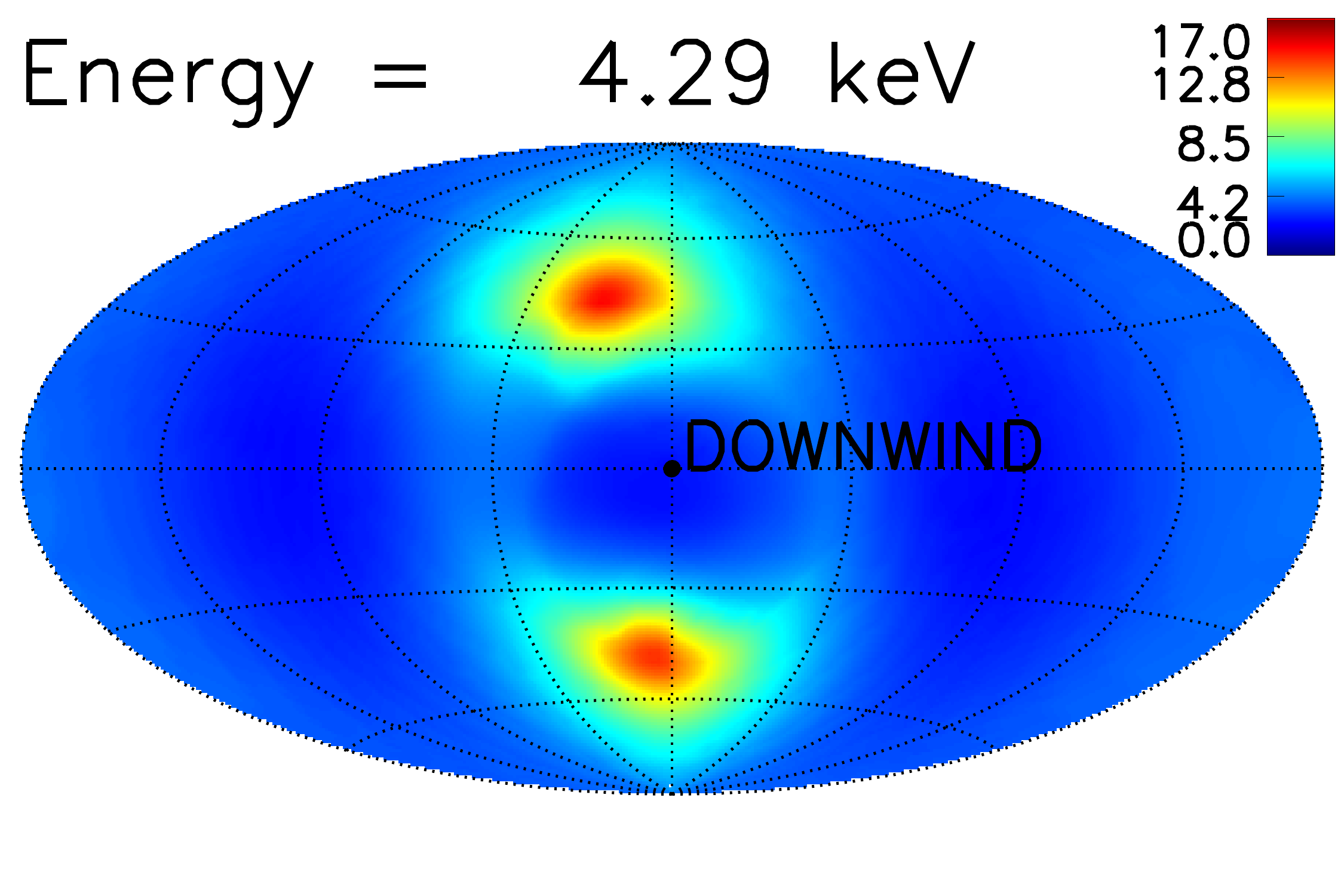}}   
 \caption{Simulated maps of ENA flux without extinction in the IHS in units of (cm$^{2}$ s sr keV)$^{-1}$. The energy
 bands of the maps are centered on (a) 1.11 keV, (b) 2.73 keV, and (c) 4.29 keV. All are 
 tail-centered maps of ENA flux.} 
\label{Maps_NoExt}
\end{figure*}

In Fig. \ref{FluxComps} we present meridional slices through the heliosphere at the 4.29 keV energy band. We show the ENA flux at each individual grid point (dJ/dr). With the presence of extinction, a majority of the flux originates close to the termination shock, with a strong contribution from the nose. Three factors which influence the signal are: transmitted PUI density (Fig. \ref{FluxComps}b), transmitted PUI temperature (Fig. \ref{FluxComps}c), and bulk PUI speed (Fig. \ref{FluxComps}d). While the ENA flux is strongly correlated with the PUI temperature, the PUI density also has a strong contribution. There are more PUIs in the nose, which contribute to the ENA flux seen in Fig. \ref{FluxComps}a. The high temperature and density at the poles, which extend backwards into the heliospheric jets, contribute to the lobes seen in  Fig. \ref{Maps}f. While extinction prevents an observer from seeing ENAs far down the tail and into the jets, the collimation of the heliospheric material seen in density and higher temperature at the poles manifests itself within the ENA maps as lobes. The PUIs maintain a higher density out to further distances in the poles as the plasma gets deflected at the nose towards the tail. This causes a higher ENA flux at the higher latitudes as compared to lower latitudes. The bulk speed of the PUIs does have an effect on the charge exchange process, but the ENA flux is not strongly correlated with this quantity because the shift in the distribution function due to the bulk speed is less effective at producing higher energy parent protons than a widening of the distribution function due to a higher temperature.

Fig. \ref{Pressure} shows the thermal pressure of the different ion components for the 4.29 keV energy band. The pressure gives insight into the influence of the different PUI populations with regard to the ENA flux as the thermal pressure includes both the density and the temperature. The primary contributors to the ENA flux are the transmitted and reflected PUIs. As is evident from Figs. \ref{Pressure}a and \ref{Pressure}b, the transmitted and reflected PUIs have a similar pressure profile. While the reflected PUI population does have a higher temperature than the transmitted PUIs, the transmitted PUIs have a similar pressure to the reflected PUIs because of their higher density. Additionally, the SW population also has a high pressure, because while the temperature is much less than either PUI population, the density is much greater for the SW population. Also, within the IHS, extinction occurs on all three ion populations originating within the supersonic solar wind. This extinction causes a drop in the pressure of the transmitted ion populations as the distance from the termination shock increases. Fig. \ref{Pressure}d shows the plasma pressure from the MHD solution in the IHS. Since the MHD solution incorporates the energy from the PUIs locally created within the IHS into the plasma, the total MHD plasma pressure does not decrease as quickly as the pressure for the included ion populations.

In Fig. \ref{Maps_NoExt}, we see the same maps as Fig. \ref{Maps}, but without extinction being included. Extinction is crucial to matching the higher flux at the nose seen in observations. When there is no extinction, the tail dominates the ENA flux signal, where the majority of the ENA flux is due to the high latitude lobes. Within the individual lobes, there is a gradient of flux which peaks near the center of the lobe. This is caused by the collimation of the plasma within the solar magnetic field lines. The thermal pressure peaks at the center of the jet collimated by the magnetic tension \citep{Drake15}. This effect is present in our maps, where the region of highest ENA flux exists where the thermal pressure peaks.

\begin{figure*}[t]
\centering
\includegraphics[scale=0.5]{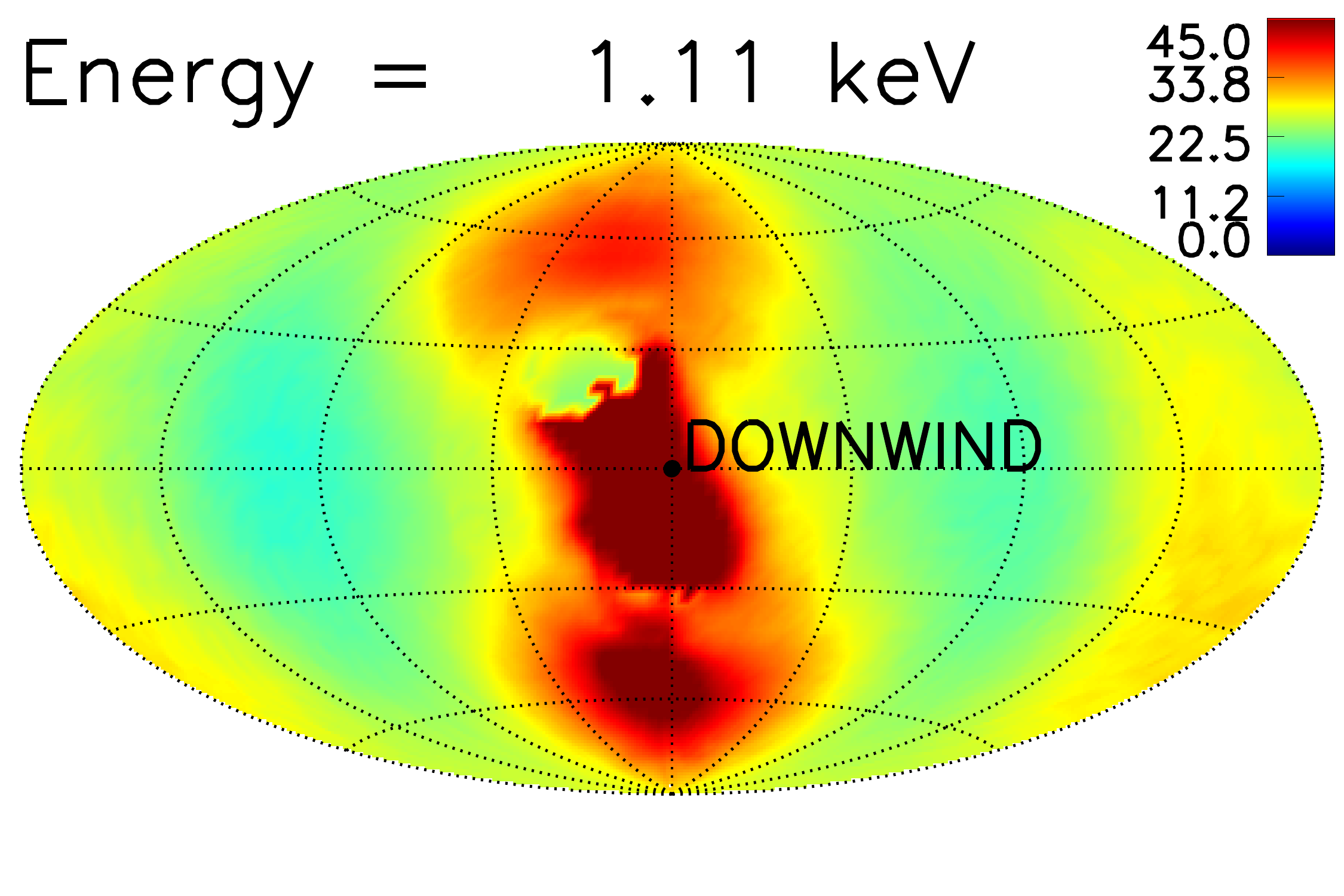}
\caption{Simulated map of ENA flux in units of (cm$^{2}$ s sr keV)$^{-1}$ for the 1.11 keV energy band when the material between the lobes of the heliosphere is incorporated into the ENA model. We assume as a proof of concept that $n_{PUI}=0.15n_{p}$ and $T_{PUI}=10T_{p}$ within this region.}
\label{Open}
\end{figure*}

\section{Discussion}
This study focuses on the effect that the newly proposed structure of the heliosphere has on ENA maps. We show that the collimation of the IHS material by the magnetic field is visible in the ENA maps. Using a uniform solar wind with a single ion plasma solution, we are able to produce high latitude lobe structures similar to those seen in both \cite{McComas13} and \cite{Schwadron14} at energies greater than ${\sim}2$ keV. It was suggested by \cite{McComas13} and by \cite{Zirnstein17} that the high latitude lobes in the tail-ward direction could be caused by the presence of fast solar wind at the poles. We find with our model that plasma collimation from the jets in the \cite{Opher15} model creates a high latitude lobe structure within the ENA maps even in the absence of slow/fast solar wind. 

We were not able to qualitatively reproduce the low latitude signal down the tail of high relative flux at lower energies ($\sim$1 keV). In the 1.1 keV map of Fig. 2, we find a weak signal down the tail, which is not present in the IBEX observations. A potential source for the strong low latitude tail signal at lower energies could be from the material between the heliospheric jets \citep{Michael18}. The PUIs in that material will be affected by the mixing between the IHS and the ISM and also potentially be accelerated due to magnetic reconnection and turbulence within the jets. As shown in \cite{Michael18}, this material also depends on magnetic dissipation. Since we do not have any information about PUIs in this region, we do not include it in our modeling, which results in a lower flux between the lobes in our simulated maps. However, in Fig. \ref{Open} we include a proof of concept for the inclusion of the material between the lobes. In this map we use our previously described method, but we also assume that 5\% of the plasma density between the lobes is composed of PUIs, and that these PUIs have a temperature of $10T_{p}$. As can be seen in this proof of concept, including the material between the lobes in our modeling significantly enhances our ENA flux in the tail at low latitudes. It is possible that \cite{Zirnstein17} do include the material that is mixed in the region of reconnection and turbulence between the jets. Exploring the effect of this material between the jets on ENA flux will be left to future work.

While we do predict lobes at high latitudes in ENA maps due to collimation, we are unable to reproduce IBEX maps on both a qualitative and quantitative level globally. Possible explanations for this are: 1) Use of a uniform solar wind model instead of a solar wind varying in latitude and time, 2) the method used to simulate PUIs, and 3) not including the material between the heliospheric jets, as noted earlier. The uniform solar wind cannot capture the variation between the fast and slow solar wind. During times of solar maximum, the uniform solar wind model can be seen as an accurate description, but during solar minimum the fast solar wind should create a hotter, less dense plasma in the polar regions of the heliosphere, which would change the results of our ENA maps. The inclusion of a time dependent solar wind with solar cycle variations would lead to an alternation between fast and slow solar wind at the poles, which would affect the ENA signal both in the polar regions near the termination shock and further back in the lobes. The solar cycle variations could also have an effect on the intensity of the solar magnetic field as shown in \cite{Michael15}, which could have an effect on the collimation as well. Additionally, since the PUI density and temperature ratios used will directly affect the ENA flux, the particular method used to include these ratios may affect the ENA flux as well. Using a denser or hotter transmitted PUI population could increase the ENA flux originating from these ions.

We do not believe our inability to reproduce IBEX maps on a qualitative or quantitative level could be explained by the heliotail resembling a comet-like shape as opposed to the ``croissant" shape we are modeling here. At the energies we are investigating, the cooling length is sufficiently short such that we cannot probe the heliopause in the low latitude tail direction in the IBEX energy range because the parent ions are depleted at this distance. While the distance to the tail at low latitudes in the ``croissant" heliosphere is considerably shorter than in the traditional comet-like structure, the distance is still larger than the cooling length for energies $>$ 20 keV.

Our inability to match with IBEX, especially at low latitudes, raises some interesting questions. As mentioned above, one possibility is that the results are based upon how the different ion populations are defined at the termination shock and throughout the IHS. Additionally, in a single ion MHD model, the populations are dependent upon how the plasma acts, yet PUIs likely do not follow the same trends (i.e. density and temperature changes) seen in the plasma solution. To properly understand this effect, a multi-ion MHD model is required to truly understand how the different ion populations act within the IHS. Additionally, when time dependence is included with solar cycle variation, we want to understand how will the high latitude lobes seen in our model be affected in terms of their ENA flux. We want to see if the inclusion of a fast solar wind, which is both hotter and less dense than the slow solar wind, would make the lobes stronger or weaker. We also find that we have a higher plasma pressure at high latitudes compared to other models (e.g. \cite{Zirnstein17}), and this higher plasma pressure can produce more ENAs at high latitudes for all energies in the range of $\sim$0.5-5 keV. Lastly, while a kinetic description of neutrals is shown to not have a strong effect on the nose of the heliosphere relative to a multi-fluid description, we want to know whether the more physical neutral solution would affect the ENA profile. These open questions will be addressed in future studies.

Within this paper we did not address the turbulence present in the lobes. In the current model, it seems that the heliospheric jets become very turbulent and experience erosion down the tail \citep{Opher15}.  We find in the current model that turbulence is strong at large distances down the tail, further than can be observed within the IBEX energy bands. This turbulence should be visible in ENA maps at 100 keV due to the increase in the cooling length at higher energies. This increase in cooling length is caused by the charge exchange profile from \cite{Lindsay05}. In Fig. 1a from \cite{Lindsay05}, the energy-dependent charge exchange cross section for a hydrogen-proton collision decreases dramatically beyond approximately 10 keV. As the extinction is based on the charge exchange cross section, a decrease in the cross section causes a decrease in charge exchange events as ions move outward from the termination shock. The cooling length at the 100 keV energy band is thus much further away from the termination shock than at the 4.29 keV band (see Fig. \ref{Stream}b). We will explore the effects of turbulence in the heliosphere on ENA maps in a future work.

\section{Summary \& Conclusions}
In the present work we show ENA maps of the ``croissant" heliosphere model. We included PUIs into our model, and allowed for both latitudinal and longitudinal variations of the PUIs. By exploring the effect of the ``croissant" heliosphere on ENA maps, we made the following conclusions:

1. \textit{Reproducing the IBEX lobes.} Using a uniform solar wind solution we were 
able to produce a high latitude lobe structure similar to those seen in IBEX ENA maps at energies greater than ${\sim}$2 keV. The presence of the lobes in our maps are caused by the collimation of the solar wind plasma via the solar wind magnetic field within the IHS, which affects the ENA flux signal even in the presence of extinction.

2. \textit{Required improvements.} We are unable to produce the strong ENA signal seen around the 1 keV energy band directly down the tail. Additionally, we are unable to quantitatively predict the flux values seen in IBEX observations. This will be explored in the future. 

\acknowledgments
The authors would like to acknowledge helpful discussions and comments from Drs. Bertalan Zieger, and Gabor 
Toth. Resouces supporting this work were provided by the NASA High-End Computing (HEC) program through the NASA Advanced Supercomputing (NAS) Division at Ames Research Center. The authors would like to thank the staff at NASA Ames Research Center for the use of the Pleiades supercomputer under the award SMD-16-7616. MO and MK acknowledge the support of NASA Grand Challenge NNX14AIBOG and NASA award NNX14AF42G.

\bibliographystyle{apalike}
\bibliography{references}

\end{document}